\documentclass{aastex63}

\received{July 22, 2020}
\revised{December 2, 2020}
\accepted{\today}
\submitjournal{AJ}

\graphicspath{{./}{figures/}}

\begin{document}

\title{Gaia-based Isochronal, Kinematic, and Spatial Analysis of the $\epsilon$ Cha Association}

\correspondingauthor{D. Annie Dickson-Vandervelde}
\email{dad1197@rit.edu}

\author{D. Annie Dickson-Vandervelde}
\affiliation{School of Physics and Astronomy, 
  Rochester Institute of Technology, Rochester NY 14623, USA}
  \affiliation{Laboratory for Multiwavelength Astrophysics, Rochester Institute of Technology}

\author{Emily C. Wilson}
\affiliation{School of Physics and Astronomy, 
  Rochester Institute of Technology, Rochester NY 14623, USA}
\affiliation{Center for Computational Relativity and Gravitation, Rochester Institute of Technology}

\author{Joel H. Kastner}
\affiliation{School of Physics and Astronomy, 
  Rochester Institute of Technology, Rochester NY 14623, USA}
 \affiliation{Laboratory for Multiwavelength Astrophysics, Rochester Institute of Technology}
\affiliation{Center for Imaging Science, 
  Rochester Institute of Technology; jhk@cis.rit.edu}

\begin{abstract}

The precise parallax, proper motion, and photometric measurements contained in Gaia Data Release 2 (DR2) offer the opportunity to re-examine the membership and ages of Nearby Young Moving Groups (NYMGs), i.e., loose groups of stars of age $\lesssim100$ Myr in the solar vicinity. Here, we analyze the available DR2 data for members and candidate members of the $\epsilon$ Cha Association ($\epsilon$CA) which, at an estimated age of $\sim$3--5 Myr, has previously been identified as among the youngest NYMGs. The several dozen confirmed members of $\epsilon$CA include MP Mus and T Cha, two of the nearest stars of roughly solar mass that are known to host primordial protoplanetary disks, and the Herbig Ae/Be star HD 104237A. We have used Gaia DR2 data to ascertain the Galactic positions and kinematics and color-magnitude diagram positions of $\epsilon$CA members and candidates, so as to reassess their membership status and thereby refine estimates of the distance, age, multiplicity, and disk fraction of the group. Our analysis yields 36 \textit{bona fide} $\epsilon$CA members, as well as 20 provisional members, including 3 new members identified here as comoving companions to previously known $\epsilon$CA stars. We determine a mean distance to $\epsilon$CA of 101.0$\pm$4.6 pc and confirm that, at an age of $5^{+3}_{-2}$ Myr, $\epsilon$CA represents the youngest stellar group within $\sim$100 pc of Earth. We identify several new photometric binary candidates, bringing the overall multiplicity fraction (MF) of $\epsilon$CA to 40\%, intermediate between the MFs of young T associations and the field. 

\end{abstract}

\keywords{Stellar associations, Pre-main sequence stars}

\section{Introduction} \label{sec:intro}

Nearby Young Moving Groups (NYMGs) serve as laboratories for the study of stars and their planetary offspring during their first stages of development, i.e., over the first tens of Myr after these young stars and planets have emerged from their birth clouds \citep{Kastner2020WP}. The age range spanned by coeval NYMGs (a few Myr to several $\times$ 100 Myr) provides unique opportunities to study stellar properties over a range of masses and temperatures at specific snapshots of pre-main sequence (pre-MS) and young MS evolution. Thanks to the proximity (${<}~ 120$ pc) of NYMGs, it is possible to observe and characterize an entire stellar population, down to the diminutive brown dwarfs \citep[e.g.,][]{Schneider2019,Phillips2020} and even massive planets \citep[e.g.,][]{Gagne2018}. The younger NYMGs (ages $<$25 Myr) feature Sun-like stars orbited by gas-rich circumstellar disks that are likely sites of ongoing planet formation \cite[e.g.,][]{Sacco2014}, while dusty debris disks are found among groups at more advanced ages \cite[e.g.,][]{Zuckerman2019}.

The precise astrometric and photometric data flowing from the Gaia Space Astrometry mission  \citep{Gaia2016,Gaia2018} is particularly useful to the study of NYMGs. Gaia Data Release 1 and the subsequent Gaia Data Release 2 (hereafter DR2) has resulted in a leap in our understanding of the membership of previously known NYMGs and, hence, their fundamental properties, such as ages, mass distributions, internal kinematics, and stellar multiplicity statistics 
 \citep[e.g.,][]{Wright2018,GagneFaherty2018,Lee2019}. Gaia DR2 data have also led to the identification of previously unknown NYMGs \cite[e.g.,][]{Gagne2018VCA}.

Here, we present an analysis of Gaia DR2 data for the $\epsilon$ Chamaeleontis Association (hereafter $\epsilon$CA). At an estimated age of 3-5 Myr \citep[][henceforth M+13]{Murphy2013}, $\epsilon$CA is the youngest NYMG within ${\sim}$100 pc of Earth \citep[][and refs.\ therein]{Kastner2020WP}. Its relative youth and consequent large disk frequency (29\%, M+13) --- combined with its position well in the foreground of the Cha star formation complex (see Fig.~2 of M+13) --- make it a key NYMG for purposes of studying planet formation and pre-MS stellar evolution, without the complication of pervasive ambient or intervening molecular cloud material. 

The identification and study of stars belonging to $\epsilon$CA began in association with the study of the neighboring $\eta$ Cha cluster \citep{Mamajek2000}. Chandra X-ray Observatory observations of the Herbig Ae/Be star HD 104237A $\epsilon$CA led to further characterization of the group and discovery of new members \citep{Feigleson2003}. HD 104237A has a close, T Tauri-type companion (HD 104237B), both of which are orbited by a CO-rich circumbinary disk \citep{Hales2014}. An additional three stars are thought to be part of the HD 104237 system, two of which form another close binary (HD 104237D+E); HD 104237D shows evidence of accreting plasma \citep{Testa2008}.

In addition to HD 104237A, the members of $\epsilon$CA include two stars with gas-rich circumstellar disks, MP Mus and T Cha; these stars represent two of only four known examples of solar-mass stars with protoplanetary disks that lie within $\sim$100 pc of Earth \citep{Kastner2010,Sacco2014}. Among these four, the disk orbiting MP Mus --- a single, 1.2 M$_\odot$ star lying a mere 98 pc from Earth --- may be most closely analogous to the solar nebula. Meanwhile the T Cha disk is the only nearby transition disk that is viewed at a high inclination (73$^\circ$), and furthermore displays evidence for an embedded planet \citep{Hendler2018}. 

The most recent comprehensive study of the membership of $\epsilon$CA (M+13) was reliant on pre-Gaia stellar kinematics and photometry. With the benefit of Gaia DR2 data, we have revisited and refined the membership, kinematics, and color-magnitude distribution of $\epsilon$CA, with the primary goal of firmly establishing its position along the NYMG age sequence. We have also reexamined the multiplicity and detailed spatial distribution of disk-bearing $\epsilon$CA members. In Section 2, we discuss the sample selection and flags and caveats of the available Gaia DR2 data. In Section 3, we describe our methods, including determination of an empirical isochrone and kinematics for the $\epsilon$CA. In Section 4, we present the results from this analysis. In Section 5, we discuss the implications of these results, including the age of $\epsilon$CA relative to other young NYMGs. 

\section{Sample Selection}
\label{sec:SampleSel}

Table~\ref{table:BOT} lists all (65) stars we considered for $\epsilon$CA membership for which Gaia DR2 data is available.  The majority of these $\epsilon$CA candidates were drawn from M+13. Those authors gathered all proposed members of $\epsilon$CA from the previous literature (52 in total) and assessed membership via their (pre-Gaia) proper motions and their spectral properties (such as Li absorption line equivalent widths and infrared excesses). The set of 35 stars initially considered here as \textit{bona fide} $\epsilon$CA members  then consists of M+13's ``confirmed members'' (M+13 Table 7).  Ultimately, we also considered the 6 ``provisional members'' and 11 ``rejected stars'' from the M+13 study (\S \ref{sec:member}) as well as 15 stars from \citet[][henceforth GF18]{GagneFaherty2018} that GF18 designated as potential $\epsilon$CA members (\S \ref{sec:othercand}).

We searched the Gaia archive for DR2 counterparts to these 67 stars from the MF+13 and GF18 studies, using a $0.25"$ search radius centered at the position listed in M+13. Unique DR2 counterparts were identified for all stars listed in Table~\ref{table:BOT} comprising a total of 65 of the 67 stars searched, though not all of these counterparts had parallaxes and colors. All Gaia counterparts met or exceeded the minimum number of Gaia visibility periods ($n_{\rm vis} > 5$) recommended by \citet{Lindegren2018}.

To assess the quality of the Gaia data for individual stars, we applied three metrics: the Re-normalized Unit Weight Error (RUWE) \citep{Lindegren2018b}, astrometric excess noise, and photometric excess \citep{Gaia2018}. The astrometric Unit Weight Error (UWE), in its renormalized form (RUWE), is useful to determine when data are potentially unreliable based on Gaia measurement uncertainties alone.  Following \citet{Lindegren2018b}, we flag data with $\mathrm{RUWE} > 1.4$.  Note that this RUWE threshold is implicitly based on a star's $G_{B_P}-G_{R_P}$ color, whereas the color-magnitude diagram analysis carried out here (Sec~\ref{sec:Isochrones}) uses Gaia $G-G_{R_P}$ colors. We also utilitzed \verb+astrometric_excess_noise+ to assess the quality of the astrometric data, flagging those stars with  \verb+astrometric_excess_noise+ $> 0.1 \times \pi$ as having potentially large parallax uncertainties. We used Gaia's \verb+phot_bp_rp_excess_factor+, $E$, to flag stars that did not satisfy $1.0 + 0.015(G_{B_P}-G_{R_P})^2 < E < 1.3 + 0.06(G_{B_P}-G_{R_P})^2$. As discussed in \citet{Gaia2018}, stars that fall outside of this range of $G_{B_P}-G_{R_P}$ have colors that may not be trustworthy.

The RUWE values for all 65 stars are listed in Table~\ref{table:BOT}. The results of the astrometric excess noise and photometric excess tests are also noted in Table~\ref{table:BOT}. Two stars, T Cha and HD 104237E, fail the \verb+phot_bp_rp_excess_factor+ test. In Figure~\ref{fig:ecplot}, we display the Gaia DR2 color-magnitude diagram (CMD) positions of the 30 (of 35) \textit{bona fide} M+13 $\epsilon$CA members that have viable DR2 parallaxes and photometry (see \S 3.1), with the results of the preceding data-flagging exercise indicated. 
\section{Analysis}

\subsection{Empirical Single-Star Isochrone}
\label{sec:Isochrones}

To ascertain the empirical isochrone defined by single-star members of NYMGs like $\epsilon$CA, we have developed the Single-star Locus Fitting Routine (SLFR, Dickson-Vandervelde et al. 2020, in prep.). The SLFR utilizes a recursive method to fit a polynomial to the color-magnitude diagram (CMD) of a set of NYMG members, identifying and rejecting outliers at each iteration. Between iterations, any star with with a magnitude greater than 2$\sigma$ is rejected before refitting the polynomial. The final result is a polynomial that represents the best fit to the single-star locus of the group across well-sampled regions of color-magnitude space, as well as a list of candidate photometric binaries, i.e., stars that lie significantly above, but within $\sim$0.75 mag of, the single-star locus. The resulting best-fit polynomial then effectively represents the group's empirical single-star isochrone.

Only 32 of the 33 \textit{bona fide} members of $\epsilon$CA listed in Table~\ref{table:BOT} have the DR2 parallaxes and photometry necessary for SLFR analysis; we further excluded two stars whose parallaxes appear incompatible with $\epsilon$CA membership (see \S~\ref{subsec:outliers}). Figure~\ref{fig:ecplot} illustrates the results of the SLFR method as applied to the remaining 30 stars. The SLFR-generated empirical isochrone (polynomial) is only well fit where there is a good sampling of stars in color-magnitude space and is particularly unreliable (and is therefore not plotted) in the bluest region of the CMD ($G - G_{RP} < 0.5$).We found a 4th-order polynomial to be sufficient to match the color-magnitude data without introducing artifacts in poorly sampled regions. The final polynomial is given by $G = 1.28c^4 + 8.44c^3 - 28.9c^2 + 33.17c + 7.18$, where $G$ is absolute $G$ magnitude and $c$ is $G - G_{RP}$ color. This fit provided an RMS in $G$ of 0.61 magnitudes. Hereafter we refer to this polynomial as the empirical (5 Myr) isochrone. 

The empirical isochrone is evidently a better match to the $\epsilon$CA color-magnitude data than any of the theoretical isochrones \citep[from][]{Tognelli2018} plotted in Figure~\ref{fig:ecplot}. While the 5 Myr and 8 Myr theoretical isochrones follow the SLFR-generated empirical isochrone in the blue CMD regions (i.e., $G - G_{RP}$ in the range 0.2--0.8), both of these isochrones fall below the NYMG distribution for redder colors ($G - G_{RP} > 0.8$). This general behavior, in which theoretical isochrones fall under the single-star loci of NYMGs in Gaia-based CMDs, has been well documented in the recent literature \citep[e.g.,][]{Gagne2018VCA}. Nonetheless, it is apparent that the 5 Myr isochrone appears to be the best overall match to the SLFR-generated empirical isochrone. We further discuss the implications of the comparison between empirical and theoretical isochrones for the age of the $\epsilon$CA in \S~\ref{subsec:ageanalysis}.

The SLFR polynomial fitting exercise yields five photometric binary candidates among the 30 \textit{bona fide} M+13 $\epsilon$CA members plotted in Figure~\ref{fig:ecplot}. Of note, the majority of these binary star candidates have high RUWE. This is consistent with the results of \citet{Belokurov2020}, who found that stars along the binary-star locus within all Gaia data show higher RUWEs than stars along the single-star locus. This correlation is likely a result of marginally resolved, close binaries yielding low-precision astrometric solutions. The SLFR method also flagged two clear CMD outliers among the 32 \textit{bona fide} members, T Cha and 2MASS J12014343-7835472, both of which are discussed in Sec~\ref{sec:retained}.

\subsection{Kinematic Analysis}
\label{sec:kinematic}

We calculated the heliocentric space motions and positions of the (35) stars considered as $\epsilon$CA members by M+13 (Section~\ref{sec:SampleSel}) for which proper motions, parallaxes, and radial velocities (RVs) are available (see Table~\ref{table:BOT}). Gaia DR2 provides RVs for 8 stars, and the remaining (majority) of the RVs come from M+13 and other RV surveys in the literature (see footnote b of Table~\ref{table:BOT}).  Heliocentric velocities ($UVW$) and positions ($XYZ$) were calculated using code from BANYAN $\Sigma$ \citep{Gagne2018BANYON}. In cases where a star had multiple RV measurements, we adopt the mean RV to calculate $UVW$, except in the case of T Cha (see Section~\ref{sec:retained}). The resulting $UVW$ were used to reevaluate $\epsilon$CA membership (\S~\ref{sec:member}).

\section{Results}

\subsection{Membership}
\label{sec:member}
\subsubsection{Kinematic and photometric inclusion/rejection parameters and criteria}

We employed two quantitative criteria, kinematic offset ($K_{kin}$) and magnitude offset ($\Delta M$), to help assess $\epsilon$CA membership. These two metrics, listed in  Table~\ref{table:deltas}, correspond to those defined in M+13. Specifically, 
the  kinematic offset is defined as
\begin{equation}
K_{kin} = \sqrt{(U-U_0)^2+(V-V_0)^2+(W-W_0)^2}
\end{equation}
where ($U_0$,$V_0$,$W_0$) is the mean space motion for $\epsilon$CA as calculated from the 30 \textit{bona fide} M+13 members used for the SLFR analysis (\S \ref{sec:Isochrones}). We defined $\Delta M$ as:
\begin{equation}
    \Delta M = M'(c) - M
\end{equation}
where $c$ is the color of the star, $M'(c)$ is the absolute magnitude of the empirical isochrone at the star's color, and $M$ is the measured absolute G magnitude of the star. Outliers in $\Delta M$ are those with $|\Delta M| \ge 2\sigma_{\Delta M}$ and in $K_{kin}$ are $K_{kin} \ge 2\sigma_{K_{kin}}$, where $\sigma_{\Delta M} = 0.629$ mag and $\sigma_{K_{kin}} = 3.2$ km s$^{-1}$. These standard deviations are calculated for the M+13 \textit{bona fide} membership list with usable data (30 stars), i.e., omitting M+13's (17) provisional and rejected members. 

Figure~\ref{fig:offsets}, like Figure 8 in M+13, illustrates the kinematic and magnitude offsets plotted against each other.  The figure includes all stars from M+13 (i.e., M+13's \textit{bona fide}, provisional, and rejected members) with Gaia parallaxes and $G - G_{RP}$ colors. Stars with $K_{kin}$ or $\Delta$M values that place them within both 2$\sigma$ boundaries, within errors, are henceforth considered to be high-probability members of $\epsilon$CA. Stars outside these bounds were reconsidered for membership on a case by case basis. For instance, the two magnitude outliers within 1$\sigma_{K}$, HD 104036 and HD 104237A, are both A-stars in the bluest region of the CMD. This region is not well fit by the empirical isochrone so the large $\Delta$M is not enough to disqualify their membership.  Additional cases, including stars whose DR2 parallaxes render their $\epsilon$CA membership doubtful, are discussed in \S 4.1.2, \S 4.1.3.

Figure~\ref{fig:provfinalcmd} presents Gaia CMDs for the full sample of (M+13) stars initially considered $\epsilon$CA members, i.e. for the stars listed in Table~\ref{table:BOT} (left panel), and for the final membership list, which is presented and discussed in \S \ref{sec:finalmembers} (right panel). The empirical isochrone in both panels is that found for the original 30 star dataset (i.e., is the same as the SLFR curve in Figure~\ref{fig:ecplot}). In the following, we describe the reasoning behind retaining or rejecting the $>2\sigma$ outliers plotted in Fig.~\ref{fig:offsets}. 

\subsubsection{Candidates rejected on the basis of kinematics, photometry, and/or parallaxes}
\label{subsec:outliers}

We corroborate the rejected status of all of the eleven $\epsilon$CA members rejected by M+13. Five of the M+13 rejects clearly maintain their non-member status on the basis of the large Gaia-based kinematic and magnitude offsets determined here (Table~\ref{table:deltas}). One rejected M+13 member with offsets $<2\sigma$, 2MASS J12074597-7816064, is unlikely to be a member given its position in the Gaia CMD (Fig.~\ref{fig:ecplot}). This star falls below our empirical isochrone, despite being a suspected spectroscopic binary (M+13). Its kinematic offset is also at the edge of acceptability ($K_{kin} = 3.03$ km s$^{-1}$). We hence exclude 2MASS J12074597-7816064 from our final $\epsilon$CA membership list. Another five M+13 rejected stars now convincingly lose membership status based on Gaia parallaxes (2MASS J11334926-7618399, RX J1243.1-7458, RX J1150.4-7704, TYC 9238-612-1, and TYC 9420-676-1). VW Cha, which was previously rejected on the basis of a pre-Gaia distance estimate, was not re-considered in this work because no Gaia parallax was reported in DR2. 

Two stars previously considered \textit{bona fide} $\epsilon$CA members by M+13, 2MASS J11432669-7804454 and RX J1220.4-7407, have CMD positions consistent with young ages and are marginally consistent with $\epsilon$CA membership (in terms of $\Delta M$) given our SLFR analysis, although both were flagged as photometric binaries due to their elevation above the single-star locus. Kinematically, RX J1220.4-7407 lies within the distribution of $\epsilon$CA members ($K_{kin} = 2.11$ km s$^{-1}$), while 2MASS J11432669-7804454 has a large kinematic offset (17.9 km s$^{-1}$). However, these stars have Gaia parallaxes of $5.54$ and $6.71$ mas, respectively, a factor of 1.5-2 smaller than the mean for $\epsilon$CA ($9.81$ mas). Notably, both stars have high RUWE and \verb+astrometric_excess_noise+ values, suggesting that their parallax solutions may not be reliable. If their parallaxes are in fact accurate, their relatively large distances would indicate they are younger, background T Tauri stars associated with the Cha star-forming region \citep[see, e.g.,][]{Kastner2012}. Their positions above the single-star locus would be consistent with their belonging to this (generally younger) Cha cloud population. Given these uncertainties, we do not include these two stars in our final roster of \textit{bona fide} $\epsilon$ CA members, although their status is worth revisiting in future Gaia data releases.

Of the six provisional members identified by M+13, two can be rejected outright based on their Gaia parallaxes: TYC 9414-191-1 ($\pi = 1.56$ mas) and CM Cha ($\pi = 5.15$ mas). TYC 9414-191-1 is likely a background star with proper motions similar to those of $\epsilon$CA members, and CM Cha (like 2MASS J11432669-7804454 and RX J1220.4-7407) appears to belong to the more distant Cha cloud T Tauri star population. The other four M+13 provisional members are discussed in Section~\ref{sec:othercand}.

\subsubsection{Outliers retained as likely members}
\label{sec:retained}

\textit{T Cha:} T Cha was previously established as a kinematic member of $\epsilon$CA \citep{Torres2008}. The kinematic analysis of T Cha is more complicated than other $\epsilon$CA members, however, due to its variable RV; measurements range from $\sim$6 to $\sim$30 km s$^{-1}$ \citep{Schisano2009}. We adopt the RV measured by \citet{Guenther2007}, 14.0$\pm$1.3, which is near the median of the values reported by \citet{Schisano2009}; with this RV, the $UVW$ of T Cha (presented in \S~\ref{sec:finalmembers}) yield a $K_{kin}$ of $3.0\pm1.3$ km s$^{-1}$, well within the 2$\sigma$ region. Adopting a larger value for its RV (e.g., the Gaia DR2 RV, 25.52$\pm$4.24) would make T Cha a kinematic outlier. However, its $XYZ$ (52.7, -90.1, -31.4 pc) positions T Cha near the median of $\epsilon$CA. We conclude that the Gaia DR2 astrometric data for T Cha support its membership in $\epsilon$CA, although the variable nature of its RV still remains to be characterized \citep[see discussion in][]{Schisano2009}.

\begin{figure}
    \centering
    \includegraphics[width=\linewidth]{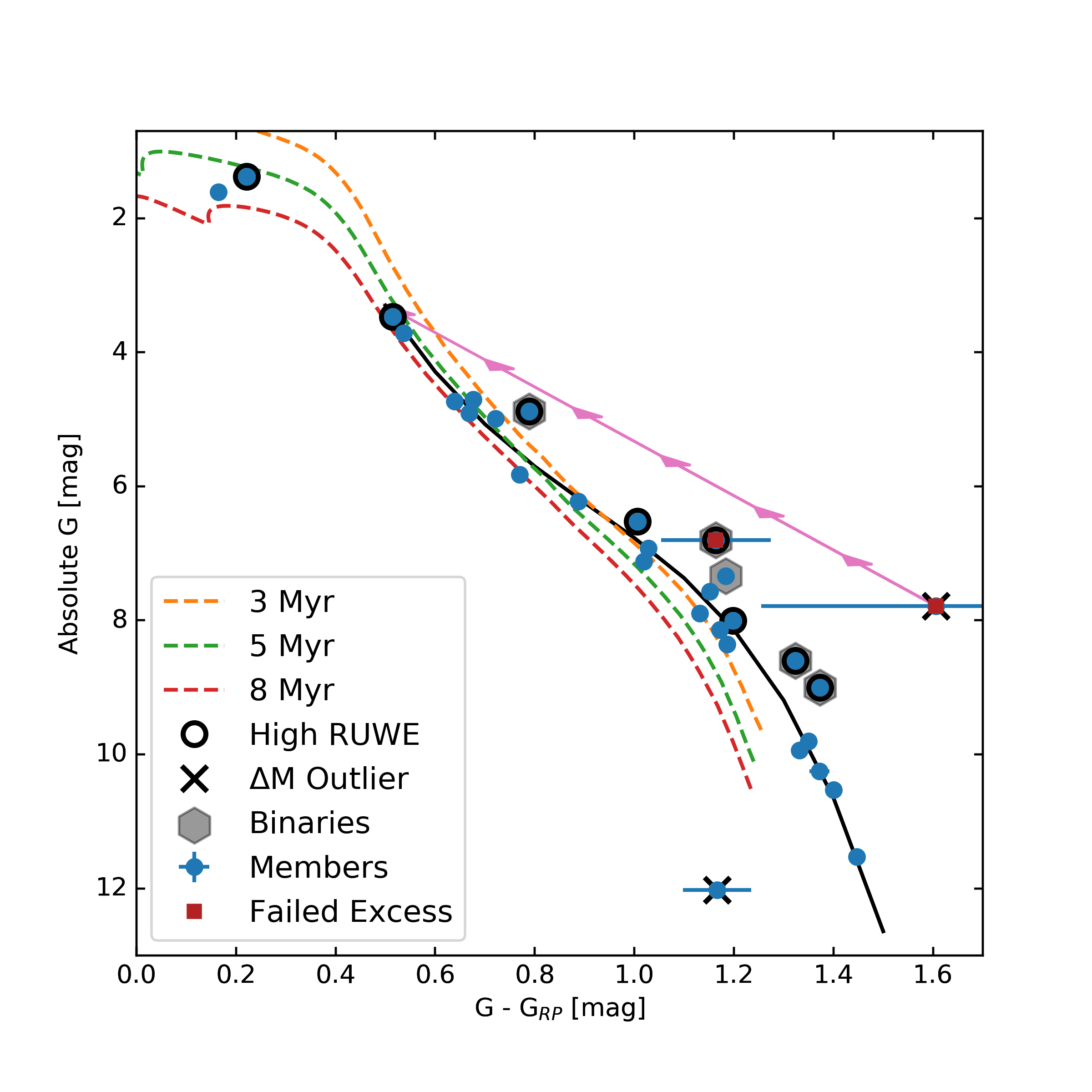}
    \caption{Gaia DR2 color-magnitude diagram (CMD) for the 30 M+13 \textit{bona fide} members (blue circles) included in empirical single-star locus fitting (SLFR) analysis (Sec~\ref{sec:Isochrones}). The best-fit empirical isochrone obtained from the SLFR method is represented by the black curve. Three theoretical isochrones \citep{Tognelli2018} for ages of 3, 5, and 8 Myr (orange, green, and red dashed lines, respectively) are also overlaid on the data. Stars with high RUWE values (low-accuracy astrometry) are marked with black, open circles; stars failing the color excess factor test (bad photometry) are marked with red squares; and stars that are outliers in the magnitude offset are marked with crosses. Five stars identified as candidate photometric binaries (via the empirical single-star isochrone fitting) are denoted by grey hexagons. Errors are displayed as horizontal and vertical bars in blue; where no error bar is seen, the errors are smaller than the symbols. The pink line (with arrows) represents the reddening vector inferred for T Cha (the red square and cross at $G-G_{RP} \sim 1.6$), i.e., $E(G-G_{RP}) = 1.1$ mag and $A_G = 4.5$ mag (see Sec~\ref{sec:retained}). }
    \label{fig:ecplot}
\end{figure}

\begin{figure}
    \centering
    \includegraphics[width=\linewidth]{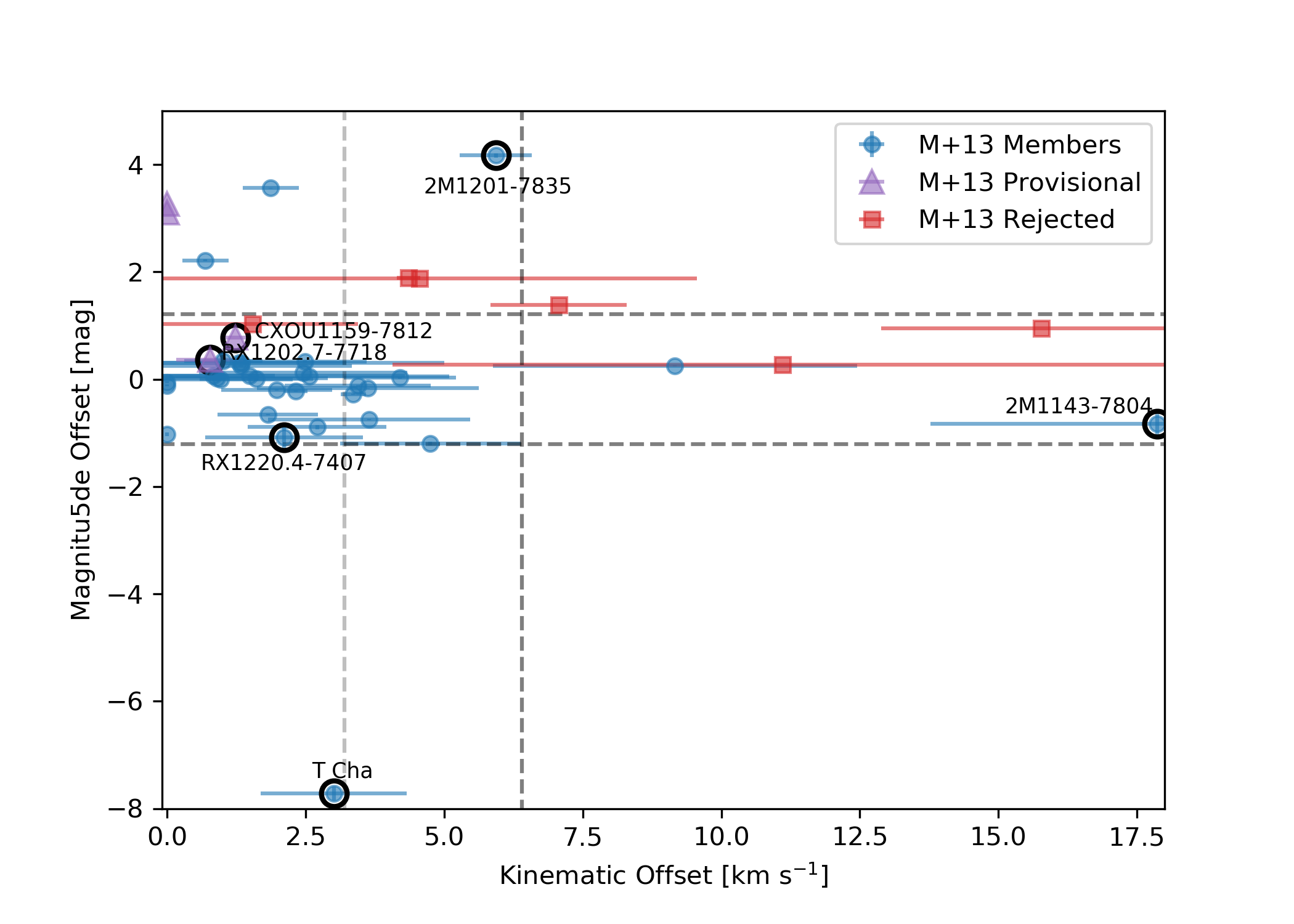}
    \caption{Comparison of the offsets in CMD ($G$ vs.\ $G-R_p$) space and kinematic ($U V W$) space used to assess $\epsilon$CA membership (Section~\ref{sec:member}). Blue circles, purple triangles, and red squares represent the M+13 $\epsilon$CA study's \textit{bona fide} members, provisional members, and rejected stars, respectively. The horizontal dashed lines correspond to $2\sigma_{\Delta M} = 1.25$ mag; the vertical dashed lines correspond to $\sigma_{K_{kin}}$ ($3.2$ km s$^{-1} $) and 2$\sigma_{K_{kin}}$. Points appearing at $0.0$ on the x-axis are stars lacking radial velocities measurements, for which kinematic offsets cannot be determined. Four outliers discussed in the text are labeled: RX J1220.4-7407 and 2MASS J11432669-7804454 in Section~\ref{subsec:outliers}, T Cha and 2MASS J12014343-7835472 in Section~\ref{sec:retained}, and RX J1202.8-7718 and CXOU J1159082-781232 in Section~\ref{sec:othercand}.}
    \label{fig:offsets}
\end{figure}

\begin{figure}
    \centering
    \includegraphics[width=\linewidth]{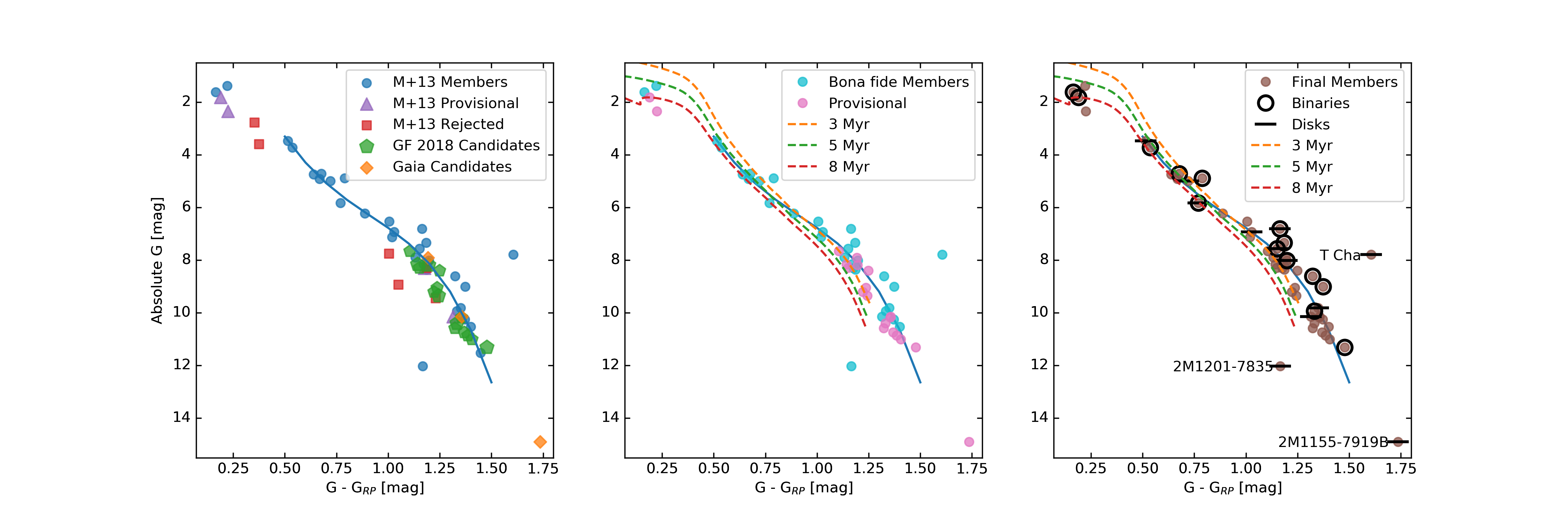}
    \caption{Gaia DR2 CMDs for the initial sample of stars considered for $\epsilon$CA membership (Table~\ref{table:BOT}; left panel) and for our final member lists, both {\it bona fide} and provisional members (Tables~\ref{table:finalmemb}, \ref{table:provmemb}); center and right panels). The empirical isochrone obtained from SLFR analysis of the original 30 {\it bona fide} members from M+13 (Fig.~\ref{fig:ecplot}) is overlaid in all three panels (blue curves). In the left panel, the blue circles, purple triangles, and red squares indicate M+13 membership candidacy (as in Fig.~\ref{fig:offsets}); new candidate $\epsilon$CA members identified via our wide-separation companion search (Sec~\ref{subsubsec:widesep}) are represented by orange diamonds; and candidates originating from GF18 are represented with green pentagons. In the center panel, our final confirmed $\epsilon$CA members (Table~\ref{table:finalmemb} are indicated with brown circles and our provisional $\epsilon$CA members (Table~\ref{table:provmemb}) are indicated with pink circles. This panel and the right panel also include theoretical isochrones (dashed curves) from \citet[][]{Tognelli2018}  for ages of 3.0 Myr (orange), 5.0 Myr (green), and 8.0 Myr (red). In the right panel, both provisional and confirmed $\epsilon$CA members (Tables~\ref{table:finalmemb} and \ref{table:provmemb}) are represented with brown circles; black dashes indicates stars with disks and black open circles indicate Gaia unresolved binaries and photometric binary candidates. Three stars of particular interest, discussed in the text, are labeled: T Cha (Sec~\ref{sec:retained}), 2MASS J12014343-7835472 (Sec~\ref{sec:retained}), and 2MASS J11550336-7919147 (Sec~\ref{subsubsec:widesep}). }
    \label{fig:provfinalcmd}
\end{figure}

The Gaia DR2 photometric data for T Cha (Fig.~\ref{fig:ecplot}) show it to be far redder and fainter than expected for an $\epsilon$CA star of its spectral type and mass \citep[G8 and 1.3 M$_\odot$, respectively;][]{Schisano2009}. In DR2, T Cha was flagged as having a suspect \verb+phot_bp_rp_excess_factor+ factor (i.e., $E=2.2$), indicative of poor-quality photometry. This is perhaps due to the star's notable variable behavior: T Cha is known to exhibit variable extinction in the optical regime, as a consequence of its highly inclined, dusty disk   \citep[$i=73^\circ$;][]{Hendler2018}. \citet{Schisano2009} found that T Cha shows a visual extinction (corrected for ISM extinction) of $\sim$0.5 magnitudes on average, but with large excursions, sometimes reaching a maximum of $\sim$3 magnitudes. Variable extinction in young stars with disks, such as that displayed by T Cha, has been hypothesized to result from a nested inner/outer disk structure wherein a warped inner disk causes quasi-periodic occultations of the photosphere \citep{Alencar2010}. Modeling of the disk around T Cha shows that instead its variability could be caused by an asymmetric, puffed-up inner disk rim \citep{Olofsson2013}. 

Based on its position in the Gaia CMD (Fig.~\ref{fig:provfinalcmd}, right) and assuming a spectral type of G8 \citep{Schisano2009} --- which suggests an absolute $G$ magnitude of 3.3, given the SLFR-derived empirical isochrone (Fig~\ref{fig:ecplot})  --- and applying a standard ISM reddening law \citep{Cardelli1989}, we infer an $A_V$ of 6.0 magnitudes at the time of Gaia observations. This $A_V$ is a factor $\sim$10 larger than the typical $A_V$ reported by \citet{Schisano2009} --- a surprising result, given that (as of DR2) Gaia data included 9 viable visibility periods. Alternatively, if we were to accept the $G-G_{RP}$ of T Cha at face value, then the star displays a color excess $E(G-G_{RP}) = 1.11$, which would imply $A_V\sim3.5$ assuming T Cha's obscuration follows the same \citep{Cardelli1989} standard reddening law. This is still a factor $\sim$5 larger than typical for T Cha, and is furthermore discrepant with $A_V$ as determined from its spectral-type-based absolute magnitude. Thus, while T Cha's anomalous position in the $\epsilon$CA CMD is likely due in part to obscuration by its disk, and suggests the disk dust exhibits non-ISM-like reddening, we caution that its position (low and red) in the CMD may also be a consequence of suspect Gaia photometry. 

\textit{2MASS J12014343-7835472:} This early-M star \citep[M2.25;][]{Luhman2004}, also known as $\epsilon$CA 11, is notably underluminous for a star in this spectral type regime ($\Delta M \sim 4$ mag; Fig.~\ref{fig:ecplot}). Previous studies of $\epsilon$CA have proposed that 2MASS J12014343-7835472 is orbited by a nearly edge-on disk \citep{Luhman2004}, such that the cold, outer parts of the disk flare to block the stellar photospheric emission, which is then only detected in the form of scattering off the inner disk \citep{Fang2013}. The star also appears to be actively accreting disk material, given its large H$\alpha$ equivalent width \citep{Luhman2004}.

Although 2MASS J12014343-7835472 has a large $K_{kin}$ value ($5.7$ km s$^{-1}$), this is mainly due to its low measured RV (11.44$\pm$2.53 km s$^{-1}$); its Gaia DR2 proper motions are generally consistent with those of other $\epsilon$CA members. 
Furthermore, its lithium absorption line ($\lambda$6708) equivalent width is consistent with the young age of the $\epsilon$CA (M+13). Its membership in $\epsilon$CA is hence supported by the available data, provided its photospheric emission is indeed strongly attenuated by a nearly edge-on disk. Unlike T Cha, the $G-G_{R_P}$ color of 2MASS J12014343-7835472 appears consistent with its spectral type (Fig.~\ref{fig:ecplot}), suggesting that the occulting disk dust has a significant component of large grains even as the Gaia photometry is dominated by starlight scattered off the disk surface.

\subsubsection{Other epsCA Candidates}
\label{sec:othercand}
\textit{Provisional M+13 candidates:} Two stars designated as provisional $\epsilon$CA members in the M+13 $\epsilon$CA study, RX J1202.8-7718 and CXOU J115908.2-781232, are upgraded here to \textit{bona fide} members. RX J1202.8-7718 was identified as a kinematic member in M+13 but its status was uncertain, with the possibility remaining that it could belong to the Lower Centaurus Crux (LCC) subgroup of the Scorpius-Centaurus OB association. In our analysis, RX J1202.8-7718 has both low $K_{kin}$ ($0.31$ km s$^{-1}$) and $\Delta M$ ($0.39$ mags), suggesting that it is unlikely to be an interloper from the LCC. Likewise, on the basis of its pre-Gaia proper motion and distance, M+13 suspected that CXOU J115908.2-781232 may be associated with the Cha I cloud. However, Gaia DR2 places the star at $\sim 105$ pc, consistent with $\epsilon$CA and much nearer to Earth than Cha I. Given its low $K_{kin}$ ($0.31$ km s$^{-1}$) and $\Delta M$ ($0.82$ mag), we include CXOU J115908.2-781232 among our \textit{bona fide} $\epsilon$CA member list. 

Two other M+13 provisional members, HD 105234 and HIP 59243, lack the requisite RVs to determine their $UVW$ and, hence, kinematic offsets, and are thus also classed as provisional members in our final membership list. Therefore, of the six provisional members presented by M+13, two are rejected, two are upgraded to \textit{bona fide}, and two maintain their provisional status. 

\begin{figure}
    \centering
    \includegraphics[width=\linewidth]{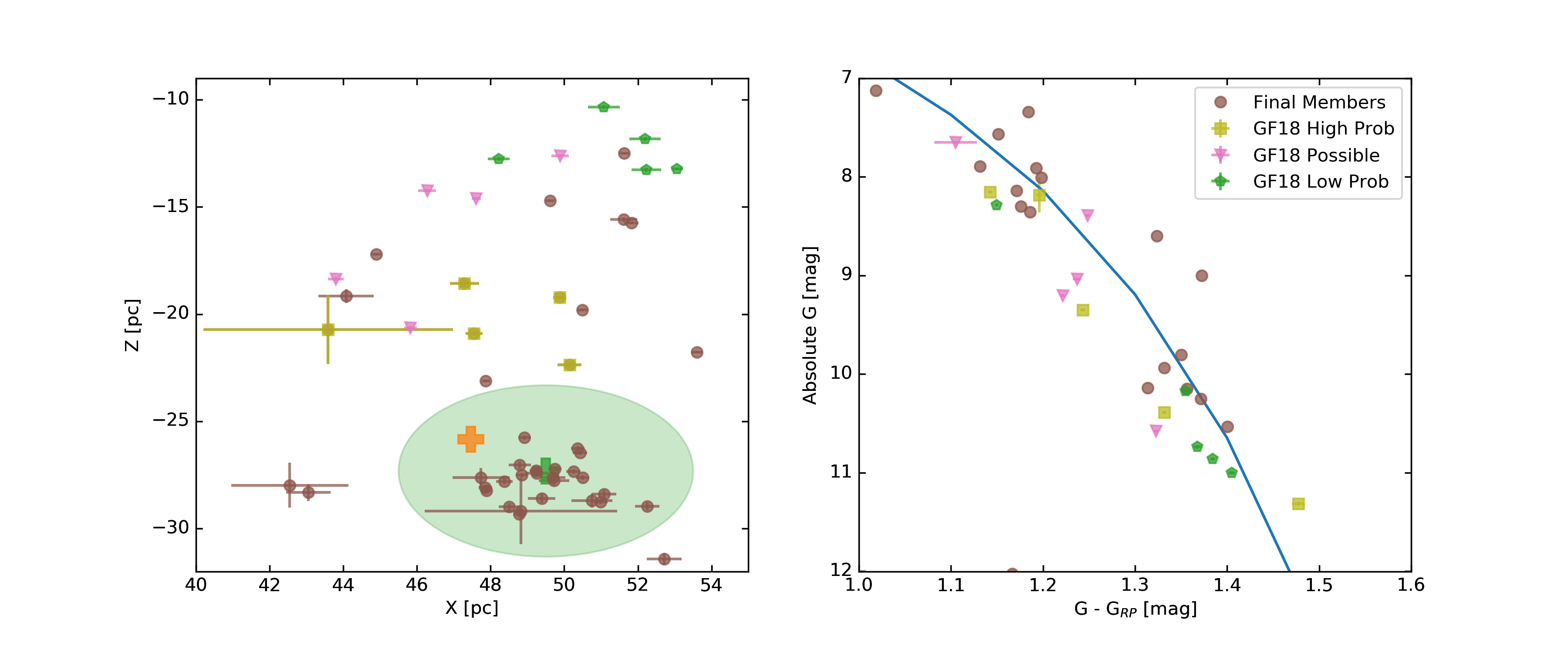}
    \caption{Individual components of heliocentric positions ($X Z$) and the Gaia CMD for both the final membership list of $\epsilon$CA (Table~\ref{table:finalmemb}; brown circles) and GF18 $\epsilon$CA candidates. Yellow squares, pink triangles, and green pentagons indicate GF18's high probability, possible, and low probability members, respectively. In the left panel, the heliocentric median and mean positions for $\epsilon$CA members (Table~\ref{table:epschameans}) are indicated as green and orange plus signs, respectively. The green shaded region represents the area within the inferred tidal radius ($\sim$4 pc; Section~\ref{sec:structure}) as centered on the median position.}
    \label{fig:gfzoom}
\end{figure}

\textit{\citet{GagneFaherty2018} candidates:} On the basis of a statistical analysis of Gaia DR2 data \citep[utilizing BANYAN;][]{Gagne2018BANYON}, GF18 identified a total of 15 stars that are candidate members of $\epsilon$CA. These stars were designated as high-probability members (5 stars), possible members (5), and low-probability members (5), based on BANYAN probabilities of membership in $\epsilon$CA as well as in other young moving groups. Their $XZ$ distribution (Figure~\ref{fig:gfzoom}, left panel) makes apparent that the GF18 high-probability members are found within the spatial locus of stars we designate as {\it bona fide} $\epsilon$CA members --- albeit outside the core $\epsilon$CA group (centered at $X \sim 49$ pc, $Z\sim -28$ pc), since the GF18 study excluded stars with distances $>$100 pc --- whereas the lower-probability members are found at the upper $XZ$ periphery of the group.

In contrast, there is no such clear distinction between the CMD positions of the three categories of GF18 candidates. All 15 stars lie along the single-star CMD locus of $\epsilon$CA determined via our SLFR analysis, although most (13 of 15) lie below the empirical isochrone, by up to a magnitude (Figure~\ref{fig:gfzoom}, right panel). While this might cast doubt on their $\epsilon$CA membership, we caution that most of the GF18 stars lie in a region of CMD space that is poorly sampled by kinematically verified $\epsilon$CA members. Hence, the final membership status of all 15 stars is not easily assessed, and must await measurements of, e.g., RVs and Li absorption line strengths.

\begin{figure}
    \centering
    \includegraphics[width=\linewidth]{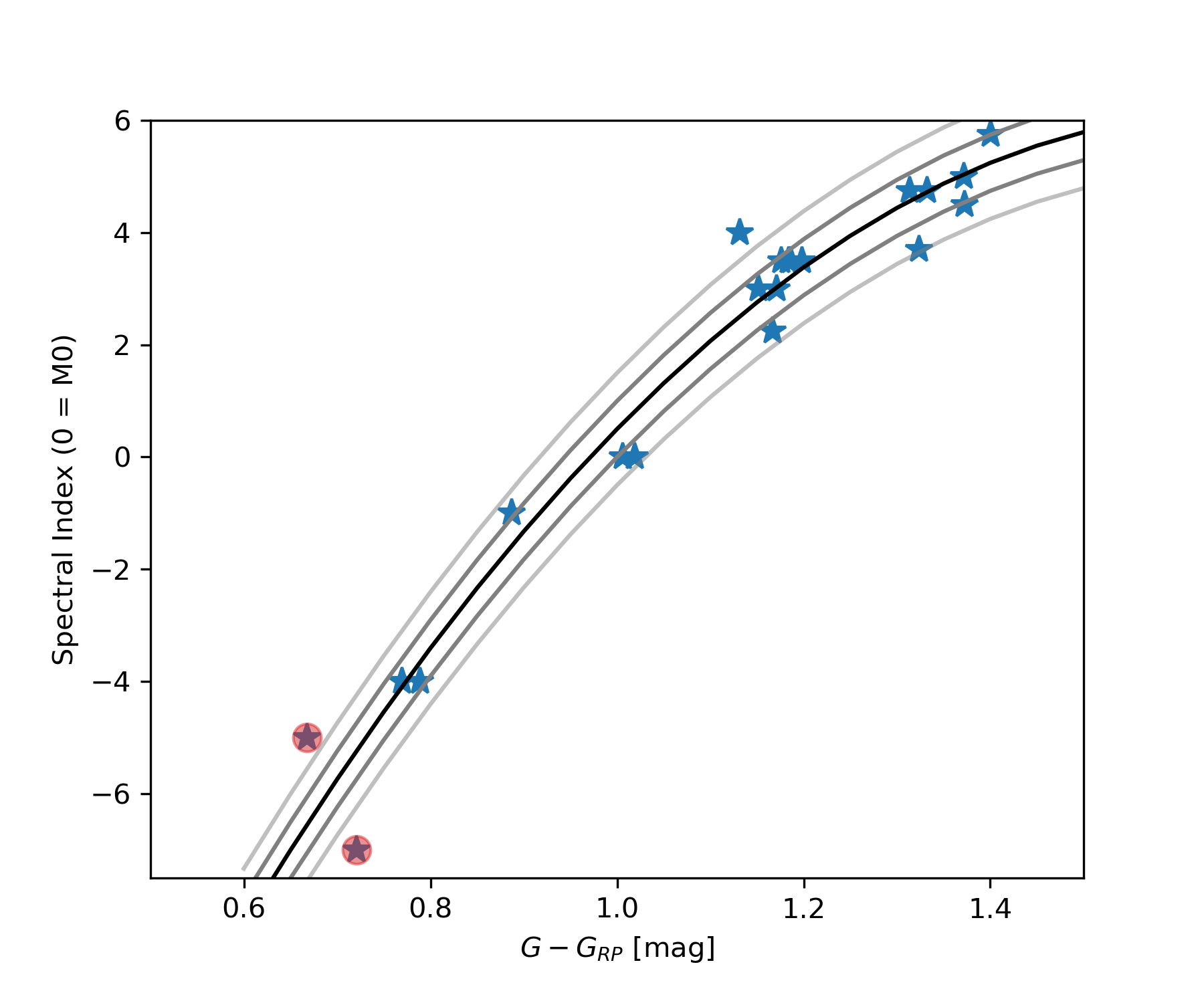}
    \caption{Spectral index  vs.\ Gaia $G-G_{RP}$ color for all M+13 members with $E(B-V) \le 0.05$. Spectral index is here defined such that (...$-1$, 0, +1, ...) = (...K7, M0, M1, ...). Most stars have inferred $E(B-V)= 0.0$; two stars with $E(B-V)=0.05$ are marked with red circles. The black curve represents the best-fit 2nd-order polynomial, and the dark and light grey curves represent deviations of 0.5 and 1.0 subtype, respectively.}
    \label{fig:SpTpoly}
\end{figure}

To assign spectral types to the provisional $\epsilon$CA members gleaned from the GF18 study, as well as other provisional members (\S \ref{sec:finalmembers}), we have determined an empirical relationship between spectral subtype and Gaia color for {\it bona-fide} K- and M-type $\epsilon$CA members. For this purpose, we selected those K and M stars from our final membership list (\S \ref{sec:finalmembers}) for which spectral types were determined from optical spectroscopy, and for which $E(B-V) \le 0.05$, based on data presented in M+13. The resulting empirical relationship is illustrated in Figure~\ref{fig:SpTpoly}. The best-fit second-order polynomial shown in the Figure is given by
\begin{equation}
    Sp(c) = -12.78c^2 + 42.54c - 29.26,
    \label{eq:sptype}
\end{equation}
where Sp is the spectral index of the star, defined such that (...$-1$, 0, +1, ...) = (...K7, M0, M1, ...), and $c$ is $G - G_{RP}$ color. Figure~\ref{fig:SpTpoly} demonstrates that Eq.~\ref{eq:sptype} provides an accuracy of roughly a subtype for K and M stars in the $\epsilon$CA. The spectral types determined from this empirical color-subtype relationship for the provisional members from GF18 are, in most cases, 0.5-2.0 subtype later than those reported by GF18. Our determinations are likely to be more reliable, however, given that they are based on the Gaia colors of {\it bona fide} $\epsilon$CA members of known spectral type, and the color-subtype relationship is known to be age-dependent \citep[e.g.,][]{Pecaut2013}.

\subsubsection{Final epsCA Membership; Provisional Members}
\label{sec:finalmembers}

Our final $\epsilon$CA membership list, including spectral types, distances, Gaia photometry, and 2MASS photometry \citep{2MASS}, is presented in Table~\ref{table:finalmemb}. Most $\epsilon$CA candidates without RVs are here designated as provisional members --- the exceptions being  three systems lacking RVs that were classified as \textit{bona fide} members by M+13, for which we find Gaia DR2-based $XYZ$ and CMD positions that are consistent with membership. The provisional members are presented in Table~\ref{table:provmemb}. The latter list includes three new members revealed via a search for wide-separation companions (Table~\ref{table:epschacands}; see Section~\ref{subsubsec:widesep}). 

For each star in Tables ~\ref{table:finalmemb} and ~\ref{table:provmemb}, we also indicate stellar multiplicity and presence/absence of evidence of a circumstellar disk. Multiplicity is further subdivided into photometric binaries (P), visual binaries from \citet{Briceno2017} (V), suspected spectroscopic binaries from M+13 (S), and candidate wide separation companion systems (C). Binarity and multiplicity within the final membership list is further discussed in Section~\ref{sec:multi}. 

Table~\ref{table:uvw} lists the heliocentric space positions ($XYZ$) and velocities ($UVW$) of the final $\epsilon$CA membership. The medians and (uncertainty-weighted) means of these spatial and kinematic coordinates, as calculated from the positions and velocities of \textit{bona fide} members in Table~\ref{table:finalmemb}, are tabulated in Table~\ref{table:epschameans}.

Figure~\ref{fig:finaluvw} illustrates the individual and mean positions and velocities.  Based on our final membership, the mean distance to the $\epsilon$CA is $100.99$ pc, with a standard deviation $\sigma = 4.62$ pc. The structure and spectral type distribution of the association are discussed in Section~\ref{sec:structure}.

\subsection{Multiplicity}
\label{sec:multi}

\subsubsection{Photometric Binaries}
\label{subsec:photo}

On the basis of our SLFR analysis, we identify five photometric binary candidates, i.e., stars that lie above the single-star locus at positions consistent with their being double or perhaps triple systems: RX J1150.9-7411, RX J1158.5-775A, HD 104237E, RX J1147.7-7842, and 2MASS J11183572-7935548. These stars have $\Delta$M values of -1.211, -0.670, -0.903, -0.767, and -1.038, respectively. Two of these systems, RX J1150.9-7411 and RX J1158.5-775A, have been resolved visually \citep[separations $0.875''$ and $0.073''$, respectively;][]{Kohler2001,Briceno2017} but evidently even the former is unresolved in Gaia DR2. 

HD 104237E is already known to have a companion, HD 104237D \citep[at separation $4.24''$;][]{Briceno2017}, and would be a hierarchical triple system if HD 104237E is confirmed as a binary (see \S \ref{sec:HD104237}).  The DR2-based binary candidacy of RX J1147.7-7842 is also novel. Both systems warrant spectroscopic followup to search for RV variability, since close binaries can have suspect Gaia DR2 astrometry \citep[e.g.,][]{Kastner2018}. We note that \citet{Briceno2017} failed to detect companions at separations as small as $\sim$0.1$''$ ($\sim$10 au) for these two stars.

\citet{Briceno2017} found that the 2MASS J11183572-7935548 system (henceforth J1118AB) is a $0.92''$ separation binary  consisting of an M4.5 primary and a lower-luminosity companion that may be related to the transition-disk nature of the object. The Gaia DR2 data confirm the angular separation ($0.9''$) and establish that this corresponds to a projected physical separation of 85 au. The parallaxes for J1118AB agree, within the errors, confirming that they constitute a physical pair. Unfortunately, the only Gaia photometry for J1118B is in the $G$ band, so we could not ascertain the CMD location of this companion. J1118A was flagged as a spectroscopic binary in M+13 and, via our SLFR method, we find it is also a candidate photometric binary. In light of the presence of the faint visual companion 2MASS 1118B \citep{Briceno2017}, it appears that 2MASS 1118AB is a possible hierarchical triple system. 

\subsubsection{Wide Separation Binaries}
\label{subsubsec:widesep}

We searched for potential wide separation companions to all $\epsilon$CA members retained after applying the criteria described in Section~\ref{sec:member}. Specifically, we searched the Gaia DR2 catalog for equidistant, comoving stars by querying the catalog for all Gaia sources within a 500'' radius of the position of each of these $\epsilon$CA members, and then reordering the resulting source list by parallax. This search radius corresponds to $\sim$50 kau, or $\sim$0.25 pc, at the mean distance to $\epsilon$CA. While a search of $\epsilon$CA member fields within this radius typically returns ${\sim}$2000 stars, once ordered by parallax, potential (equidistant) companions to the star originally queried rise to the top of the list and are hence conspicuous. All stars so identified have parallaxes and proper motions within a few percent of the star searched and so were accepted as wide separation companion candidates. 

Applying this method, we have identified three new candidate members of $\epsilon$CA. Gaia DR2 data for these candidates are listed in Table~\ref{table:epschacands}. The faintest and reddest of these three, 2MASS J11550336-7919147, is described below. The other two comoving companion candidates, 2MASS J12011981-7859057 and 2MASS J12115619-7108143, are mid-M stars that fall along the empirical single-star isochrone, with $\Delta$M$\le 2\sigma_{\Delta M}$. These two  candidate comoving companions have projected physical separations from their primaries of 5.7 kau and 17.7 kau, respectively. We estimated their spectral types (M5 and M3, respectively) from the empirical relationship between Gaia $G-G_{RP}$ color and spectral type for $\epsilon$CA members described in  \S~\ref{sec:othercand}. Neither star displays evidence of a dusty disk in the form of an IR excess, i.e.,  both have 2MASS/WISE colors consistent with those of ``diskless'' young M stars of similar spectral type \citep{Pecaut2013}.

This comoving companion search also recovered multiple stars already considered $\epsilon$CA members as components of possible wide binary systems. The RX J1158.5-7754 system was matched with HD 104036. Three stars, 2MASS J12005517-7820296, CXOU J120152.8-781840, and USNO-B 120144.7-781926, were identified as a three-component comoving system. We also identified CXOU J115908.2-781232, a provisional M+13  member, as another possible component of the HD104237 multiple system (see \S \ref{sec:HD104237}).

{\it 2MASS J11550336-7919147:} The third, newly-identified wide-comoving companion, 2MASS J11550336-7919147 (henceforth 2MASS J1155-7919B), at an absolute $G$ magnitude of $\sim$15 and $G-G_{RP} \sim 1.75$, is both the faintest and the reddest object thus far identified in $\epsilon$CA \citep{DickVand2020}. As described in \citet{DickVand2020}, we find 2MASS J1155-7919B is comoving with 2MASS 11550485-7919108 (hereafter 2MASS J1155-7919A). The star 2MASS J1155-7919A was itself initially thought to be a wide-separation comoving companion to T Cha \cite{Kastner2012}, before Gaia DR2 data established that T Cha and J1155-7919A are neither equidistant nor precisely comoving \citep{Kastner2018}. The 2MASS J1155-7919AB pair has a projected separation of 5.75'', corresponding to a projected physical separation of 566 AU. In \citet{DickVand2020}, we suggested that the position of 2MASS J1155-7919B at the extreme faint, red end of the single-star locus of the $\epsilon$CA CMD reflected its likely status as a substellar object, with a bolometric luminosity ($\log{L_{bol}/L_\odot} = -3.2$) that would imply its mass is a mere $\sim$10 $M_{Jup}$. This would make 2MASS J1155-7919B the lowest-mass $\epsilon$CA member presently known --- even less massive than WISEA J120037.79-784508.3, a recently identified brown dwarf candidate and possible $\epsilon$CA member \citep{Schutte2020}\footnote{WISEA J120037.79-784508.3, which has an absolute G magnitude of 11.3, is not included in Table~\ref{table:provmemb}. } 
However, a potential alternative model to explain the large absolute $G$ magnitude and red color of 2MASS J1155-7919B is that the object is in fact a mid-M star --- possibly a near-twin to host 2MASS J1155-7919A --- that is obscured by a large column density of gray dust in a nearly edge-on disk. Under this interpretation, 2MASS J1155-7919B would be analogous to 2MASS J12014343-7835472 (= $\epsilon$CA 11; \ref{sec:retained}), but even more highly obscured by its disk. Such an alternative explanation is motivated by the fact that 2MASS J1155-7919B is consistently 5 magnitudes dimmer than its host in G, B$_p$, R$_p$, J and H (Table~\ref{table:epschacands}), and that its spectral type, as obtained from its color (via the relationship in Fig.~\ref{fig:SpTpoly}), is M6.  In a forthcoming paper (Dickson-Vandervelde et al., in prep), we further explore these these two possible scenarios for the nature of 2MASS J1155-7919B.

\subsubsection{HD 104237}
\label{sec:HD104237}

HD 104237 is a proposed quintuplet system within $\epsilon$CA, consisting of a triple system dominated by the bright Herbig Ae/Be star HD 104237A and including close binary HD 104237DE \citep{Feigleson2003,Grady2004}. Our wide-separation companion search (Sec.~\ref{subsubsec:widesep}) also suggests that previously identified $\epsilon$CA member CXOU J115908.2-781232 is a possible additional companion to HD 104237A. We also flagged HD 104237E as a possible photometric binary, as noted in \S~\ref{subsec:photo}. While stellar components A, D, and E were all resolved within Gaia DR2, B and C were not, most likely being incorporated into the Gaia point spread function of HD 104237A. 

The majority of stars in the HD 104237 system lie near the spatial median of $\epsilon$CA, the densest region of $\epsilon$CA, and the system is well within the group's tidal radius (Sec~\ref{sec:structure}). However, the Gaia DR2 parallaxes of HD 104237D and E place these two stars 6--7 pc closer to Earth than HD 104237A. Indeed, during our wide-separation companion search, HD104237D+E flagged each other, while HD104237A only matched (in terms of parallax and proper motion tolerances) with the star CXOU J115908.2-781232. If HD 104237A+E is in fact a bound system, this would imply that the Gaia DR2 parallaxes for components D and E are spurious. On the other hand, as in the case of T Cha and J1155-7919A, the star originally designated T Cha B \citep{Kastner2012,Kastner2018}, it is possible that the apparent D  and E components of the HD 104237 system are in fact merely $\epsilon$CA members that are seen in projection near the bright primary star. This would not be surprising, given the system's position near the median  $XYZ$ of $\epsilon$CA. If the potential new components of HD 104237, CXOU J115908.2-781232 and HD104237Eb, are, in fact, bound to HD 104237A, this could be a seven-star system. Alternatively, the bound components may in fact consist of a triple system comprised of HD104237Ea+Eb+D plus a quadruple system comprised of HD104237A+B+C and CXOU J115908.2-781232. Gaia DR3 should help resolve some of this uncertainty concerning the composition of the HD 104237 system, by confirming and/or refining the DR2 parallaxes to its individual components.

\section{Discussion}

\subsection{Multiplicity Fraction and Spectral Type Distribution}
\label{sec:sptypes}
In light of the preceding, we can now revisit the $\epsilon$CA multiplicity fraction (MF), or the number of systems consisting of more than one star. Previous to this work, it was found that higher mass stars of $\epsilon$CA all have companions, while the lower mass stars have a low companion frequency \citep{Briceno2017}. The binary fraction, including suspected binaries, was reported in M+13 as $36^{+10}_{-8}$\%. The MF of our final membership list (Table~\ref{table:finalmemb}) is 40\%, with 12 of 30 systems in multiple systems, which is consistent within the errors with the MF determined by M+13. 

We further separate the MF of $\epsilon$CA into mass groups, bearing in mind small number statistics. For the population of intermediate mass (A and B, n=5), solar-type (F, G, and K, n=11), and low mass (M, n=19) star systems in $\epsilon$CA we find $\rm MF=100\%$, $\rm MF=36\%$, and $\rm MF=42\%$, respectively. This can be compared with the MFs of intermediate-mass, solar-type mass, and low mass stars on the main sequence, for which the multiplicity fractions are on average $\rm MF\ge50\%$, $\rm MF=44\pm2\%$, and $\rm MF=26\pm3\%$, respectively \citep{Duchene2013}. In contrast, the MF of T associations lies in the range $\sim$66-75\% \citep{Duchene2013}. The MF of $\epsilon$CA is hence less than that of T associations and greater than that of main-sequence stars. This is consistent with the $\sim$5 Myr age of $\epsilon$CA (\S~\ref{subsec:ageanalysis}), which is intermediate between T associations and the main-sequence field. The $\epsilon$CA  hence appears to be a key group in which to study the dissolution of young binary systems as T associations evolve toward the field population. 

Figure~\ref{fig:spthist} presents the spectral type distribution obtained from our final membership list (Table~\ref{table:finalmemb}). For purposes of this histogram, stars whose published spectral types span a range (e.g., ``M3--5'') are assumed to have a spectral type in the middle of that range, and K stars of all subtypes are grouped together. The Figure demonstrates that the $\epsilon$CA spectral type distribution, which peaks at M4, resembles that of the $\chi^1$ For cluster \citep[see Figure 6 in][]{Zuckerman2019} and other nearby associations like TWA, BPMG, and Columba, many of which have far better statistics \citep[see, e.g., Figure 9 in][]{Lee2019}.

\subsection{Structure of the Association}
\label{sec:structure}

With accurate heliocentric positions, we can analyze the structure of the $\epsilon$CA NYMG in the Galactic context. Following \citet[][]{Zuckerman2019}, the tidal radius of the group can be roughly estimated as $r=R(M_c/3M_g)^{1/3}$ \citep[][]{King1962}, where $R$ is the distance between Earth and the Galactic Center ($\sim$ 8200 pc), $M_c$ is the mass of the stellar group, and $M_g$ is the Galactic mass interior to the Sun ($\sim10^{11}$ $M_\odot$). In order to obtain a rough estimate for $M_c$, we adopt masses of 3.0, 2.0, 1.0, 0.7, and 0.3 $M_\odot$ for B, A, G, K, and M stars, respectively. We then obtain an estimate for the group mass, based on the spectral types of individual {\it bona fide} members of the Association (Table~\ref{table:finalmemb}), of $M_c \sim 28$ $M_\odot$\footnote{Note that this estimate neglects the potential contribution to $M_c$ of molecular gas possibly associated with $\epsilon$CA --- or at least with T Cha --- as evidenced by detection of interstellar CO at coincident RV \citep[][]{Sacco2014}.}. Adopting this value of $M_c$, the tidal shredding radius of the group is found to be 4 pc, with a large uncertainty (given the large uncertainties in the estimates of the $\epsilon$CA and Galactic masses). Given the estimated group mass and this tidal radius, and assuming (for simplicity) a spherical stellar distribution, the mass density is ${\sim}0.13$ $M_\odot$ pc$^{-3}$ and the stellar density is ${\sim}0.18$ pc$^{-3}$. This mass density is within a factor of 2 of the estimated local Galactic disk density \citep[$\sim0.1M_\odot$ pc$^{-3}$;][]{Mamajek2016}, reflecting the fact that $\epsilon$CA is a diffuse association (as opposed to cluster). Inclusion of provisional members would increase the Association mass estimate to 36$M_\odot$ but negligibly change the aforementioned tidal radius and density calculations.

The spatial and kinematic structure of the final membership of the $\epsilon$CA is illustrated in Figure~\ref{fig:finaluvw}. The vast majority of group members ($\sim$80\%) fall within the tidal shredding radius of the median group $XYZ$ position, shown as the green shaded circle and red cross, respectively. The binaries in the group fall both inside and outside of the central region of the association, with no obvious correlation with position. 
Figure~\ref{fig:diskdist} illustrates the disk fraction versus distance from the median $XYZ$ position of $\epsilon$CA. The figure indicates that 9 of the 11 $\epsilon$CA systems known to host disks lie within 5 pc of the median position, a volume roughly coincident with that defined by the group's tidal shredding radius; the disk fraction falls from $\sim$50\% within 5 pc of the median position to $\sim$15\% beyond 5 pc of the median. This centrally concentrated distribution suggests that stars in the core region of the $\epsilon$CA are more likely to retain dusty disks, hinting at the possibility that the nascent planetary systems orbiting stars in this region are subject to more frequent and/or more catastrophic dust-generating collisions. The distribution in Figure~\ref{fig:diskdist} stands in contrast to that of the (older) $\chi^1$ For cluster (age $\sim$40 Myr), for which the majority of stars with IR excesses (hence dusty disks) lie in a narrow, spherical annulus just outside the tidal shredding radius of the cluster \citep{Zuckerman2019}. However, as noted by those authors, $\chi^1$ For also represents a particularly striking and unusual case of a large disk frequency in a $\sim$40 Myr-old cluster.

\subsection{Age of the Association}
\label{subsec:ageanalysis}

The SLFR analysis applied to $\epsilon$CA (\S \ref{sec:Isochrones}) was additionally performed for two other NYMGs, the TW Hya Association (TWA) and $\beta$ Pic Moving Group (BPMG). These NYMGs are both slightly older than $\epsilon$CA, at 8 Myr \citep{Donaldson2016} and 24 Myr \citep{Bell2015} respectively, and hence provide good references for the relative age of $\epsilon$CA as determined from their respective SLFR-based empirical isochrones. For our SLFR analysis, we used lists of stars considered \textit{bona fide} members of each group \citep{Lee2019}, totaling 30 stars for TWA and 113 stars for BPMG.

The empirical isochrones for the three NYMGs are compared in the right panel of Figure~\ref{fig:compiso}. With the exception of BPMG, the blue ends of the empirical isochrones are not well fit, due to the small sample sizes in these regions, and are hence not plotted in Figure~\ref{fig:compiso} (and are excluded from this discussion). All three of the empirical isochrones fit their respective NYMG data well in the regions $0.2 \le G - R_p \le 1.2$. In this CMD region, the empirical isochrones show the expected hierarchical pattern --- i.e., with $\epsilon$CA highest, BMPG lowest, and TWA intermediate between the two --- reflecting the relative ages of these three NYMGs. At redder colors, all three groups drop off in population and appear to display a larger degree of scatter, and hence the empirical isochrones (4th-order polynomials) do not provide as good a fit in these regions.

Three representative isochrones from \citet{Tognelli2018} are shown in the right panel of Figure~\ref{fig:compiso}\footnote{We found MIST isochrones \citep{Dotter2016, Choi2016, Paxton2011, Paxton2013, Paxton2015} to be indistinguishable from the \citet{Tognelli2018} curves, so only the latter are used as representative of the expected temporal behavior.}. While the theoretical isochrones well match the Gaia NYMG data blueward of $G - R_p \sim 0.8$ (apart from the TWA, which lacks stars in this region), these curves fall below the Gaia data at regions redder than $\simeq$0.9 in $G - R_p$. This divergence between theoretical isochrones and data is commonly observed in Gaia CMDs, as previously noted (\S~\ref{sec:Isochrones}), and can likely be ascribed to the high levels of magnetic activity of late K and M stars \citep[e.g.,][]{Gagne2018VCA}. 

Based on the close correspondence of the theoretical 5 Myr isochrone and the empirical (SLFR-generated) single-star isochrone in the region $0.5 \le G - R_p \le 1.0$, we estimate an age of $5^{+3}_{-2}$ Myr  for $\epsilon$CA, where the range of uncertainty is based on the vertical offset of the 3 and 8 Myr theoretical isochrones in this same domain of $G-R_p$. This is consistent with the age range determined via theoretical isochrone analysis in the (pre-Gaia) M+13 study, i.e., a median of 3--5 $M_\odot$ for lower-mass stars (albeit with somewhat older inferred ages for solar-mass stars), as well as with the age obtained from a reanalysis of the M+13 data by \citep[$3.7^{+4.6}_{-1.4}$ Myr;][]{Schutte2020}. Future age estimates for the $\epsilon$CA should be informed by spectroscopic observations designed to confirm the membership status of provisional members  (Table~\ref{table:provmemb}), a DR3-based search for additional members, and application of isochrones obtained from ``magnetic'' pre-main sequence evolution models \citep[e.g.,][]{Simon2019}.

\section{Summary}

\begin{figure}
    \centering
    \includegraphics[width=\linewidth]{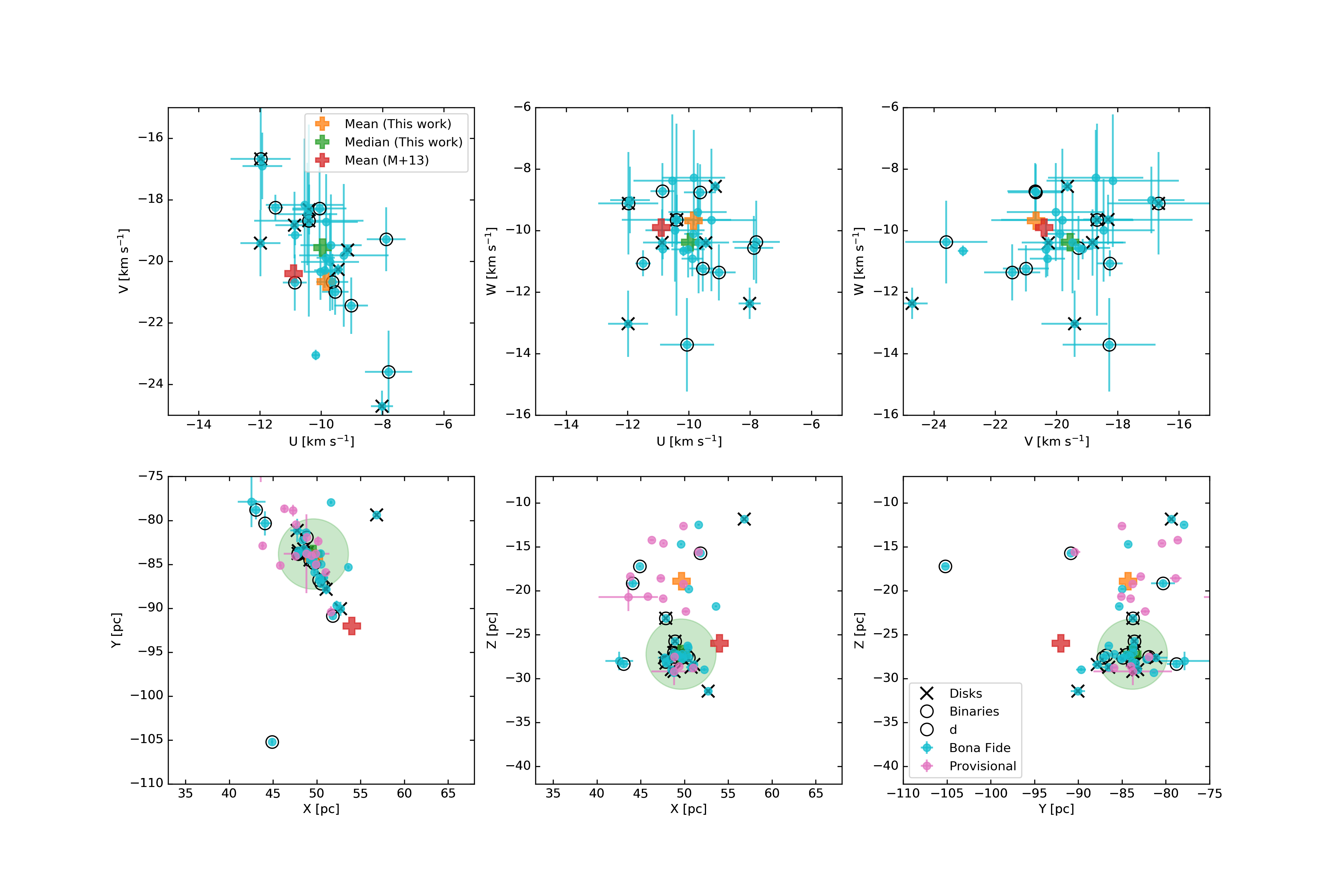}
    \caption{Individual components of heliocentric positions ($X Y Z$) and velocities ($U V W$) plotted against one another for the final membership list (Table~\ref{table:finalmemb}, cyan circles) and the provisional members (Tables~\ref{table:provmemb}, pink circles), with open black circles marking stars that are unresolved binaries and black x's marking stars with disks. Table~\ref{table:finalmemb} stars lacking measured RVs (see Table~\ref{table:BOT}) are omitted from the $U V W$ plots. The mean and median values of $U V W$ and $X Y Z$ are indicated by orange and green crosses, respectively, in each plot; the previously obtained mean values (from M+13) are represented by red crosses. The region within the calculated tidal radius (4 pc, Sec~\ref{sec:structure}) is indicated as a green circle in the $X Y Z$ plots, centered on the median. When no error bar can be seen, the error lies within the marker.}
    \label{fig:finaluvw}
\end{figure}

\begin{figure}
    \centering
    \includegraphics[width=\linewidth]{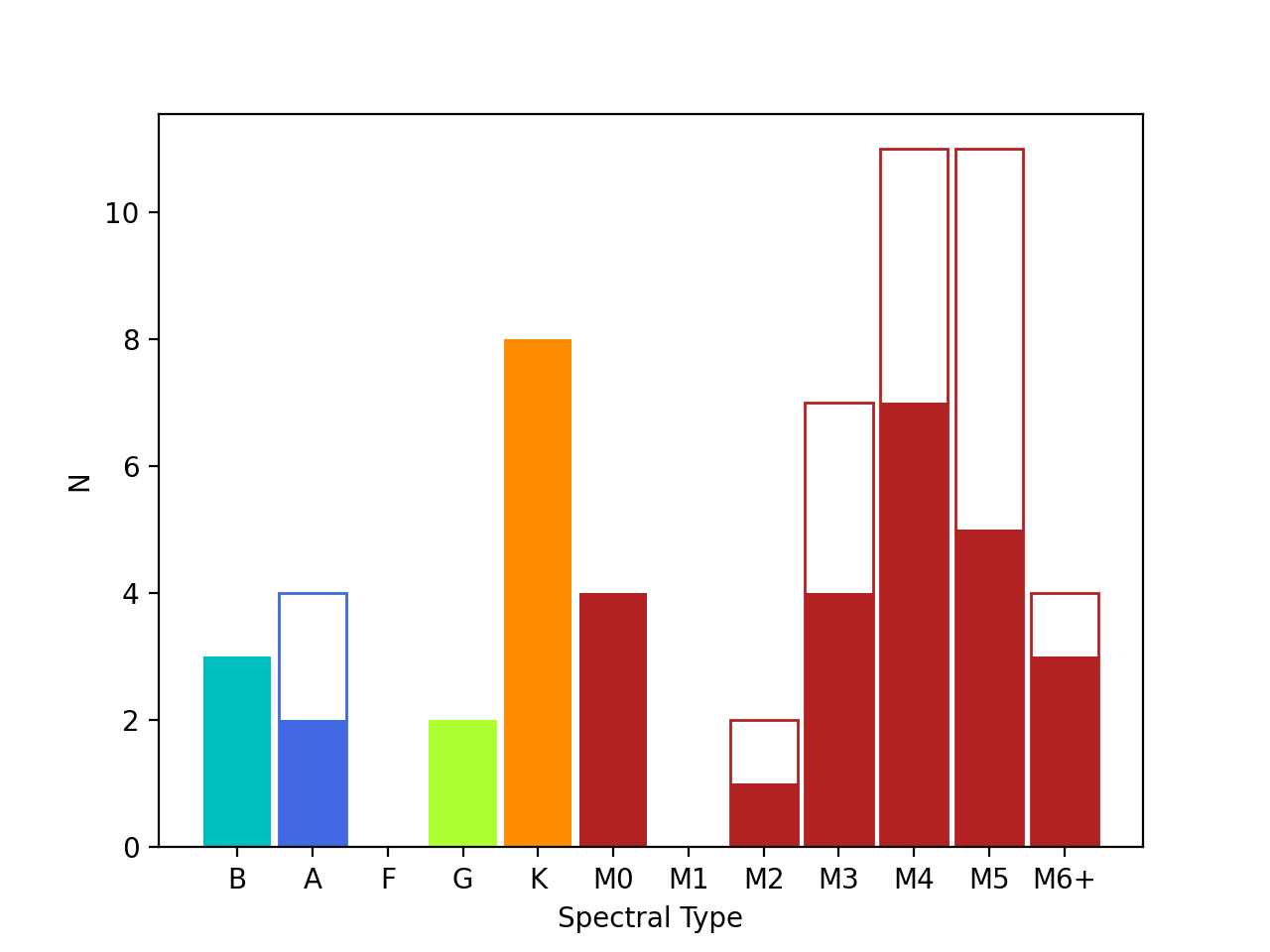}
    \caption{Spectral type histogram compiled from the final $\epsilon$CA membership lists (Table~\ref{table:finalmemb}, ~\ref{table:provmemb}). Unfilled portions of histogram bars indicate provisional members. Non-M spectral types are grouped across subtypes, while the M subtypes are plotted individually.  The color scheme follows that of Figure 9 of \citet{Lee2019}.}
    \label{fig:spthist}
\end{figure}

We have used Gaia DR2 astrometric and photometric data to refine the membership of $\epsilon$CA, and have thereby established the distance (mean $D = 100.99$ pc, $\sigma_D = 4.62$ pc), age ($5^{+3}_{-2}$ Myr), and spatial and kinematic distributions (Fig~\ref{fig:finaluvw}; Table~\ref{table:epschameans}) of the association. We confirm that $\epsilon$CA is significantly younger than both the TW Hya Association and $\beta$ Pic Moving Group (Fig~\ref{fig:compiso}) and, hence, that $\epsilon$CA represents the youngest NYMG within $\sim$100 pc of Earth. Our analysis includes the determination of an empirical relationship between Gaia $G - G_{RP}$ color and spectral type that should provide an accurate means to determine the spectral subtypes of $\sim$5 Myr-old K and M stars from dereddened Gaia photometry.

Our analysis yields a final $\epsilon$CA membership list consisting of 36 \textit{bona fide} members and 20 provisional members (Table~\ref{table:finalmemb}; Table~\ref{table:provmemb}). The provisional members require followup study (in particular, RV determinations) in order to confirm their membership status kinematically. 
These include 3 new members of the $\epsilon$CA that we have identified via a search of Gaia DR2 for wide-separation, co-moving companions to previously identified members. One of these newly identified members is either a substellar object or an M star viewed through (and hence highly obscured by) a nearly edge-on circumstellar disk. We identified 5 photometric binaries among the group members, 3 of which are new binary star candidates.

\begin{figure}
    \centering
    \includegraphics[width=\linewidth]{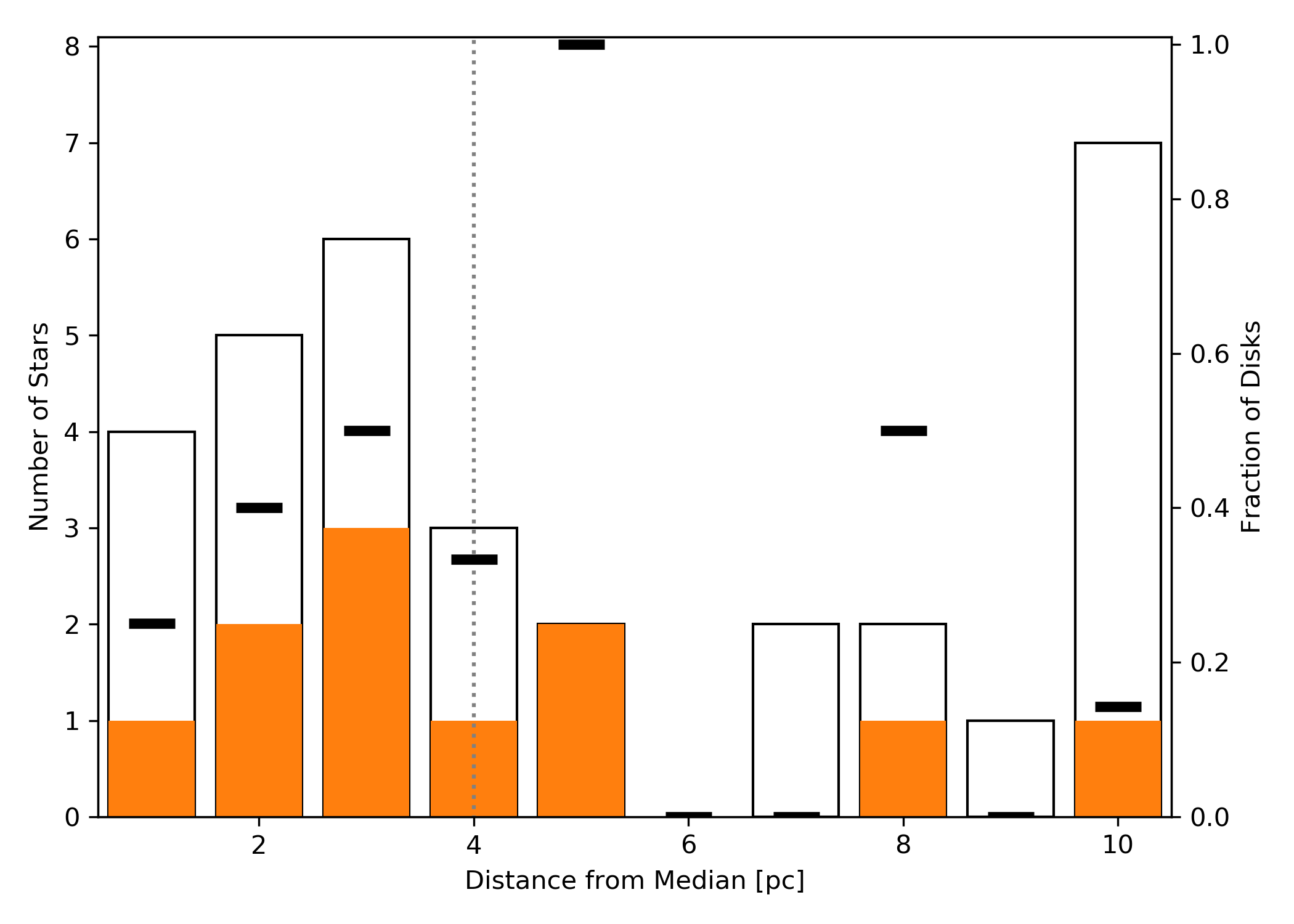}
    \caption{Histogram of the distance from the median $X Y Z$ position for the final list of $\epsilon$CA members (Table~\ref{table:finalmemb}). The filled orange bars indicate stars with disks, and the disk fractions are represented as horizontal black lines. The tidal shredding radius (4 pc; Sec~\ref{sec:structure}) is denoted by a vertical gray dashed line.}
    \label{fig:diskdist}
\end{figure}

\begin{figure}
    \centering
    \includegraphics[width=\linewidth]{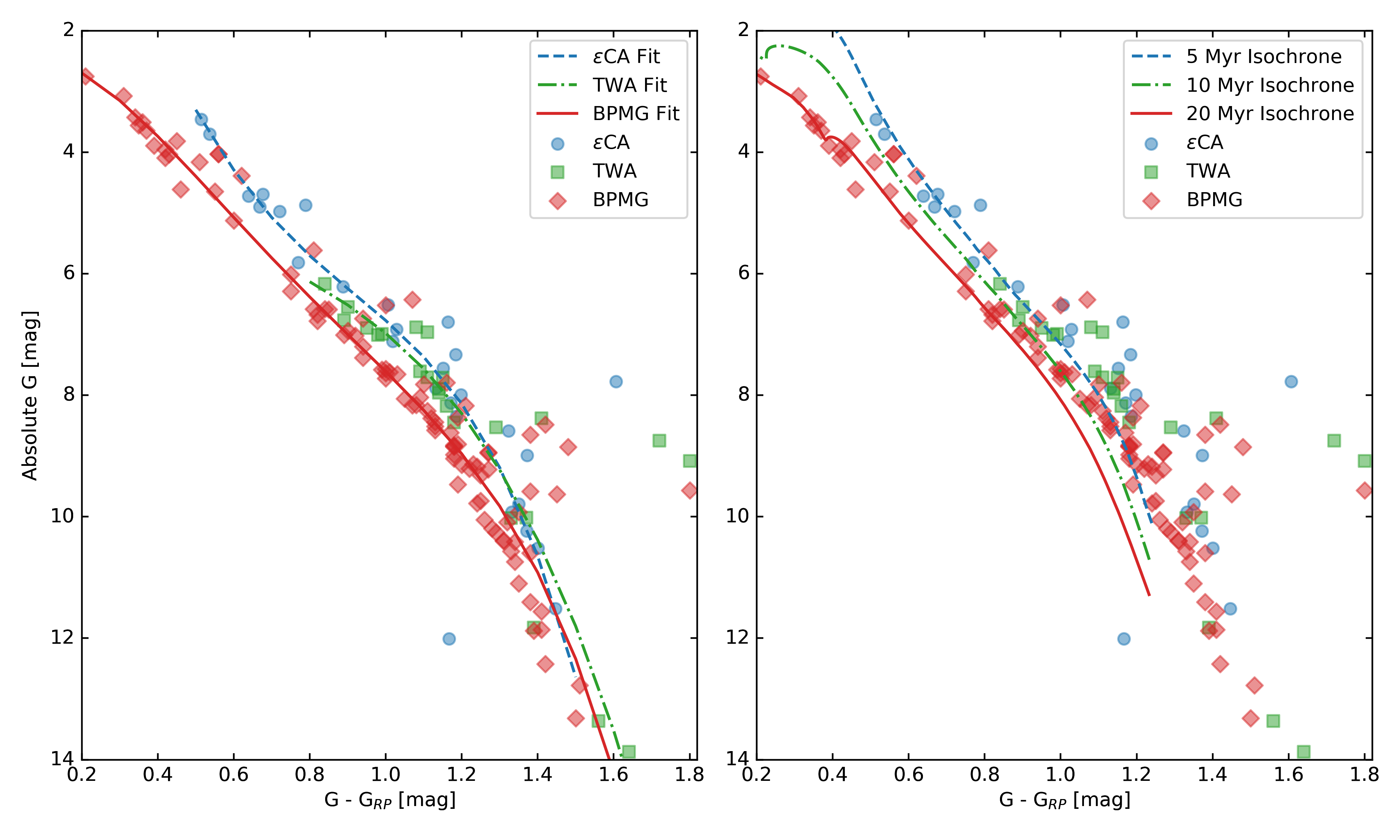}
    \caption{Gaia DR2 CMDs for $\epsilon$CA (Table~\ref{table:finalmemb}; blue circles), the TW Hya Association (green squares), and the $\beta$ Pic Moving Group (red diamonds), where data for the latter two groups are based on for the membership lists in \citet{Lee2019}. In the left panel, the data for the three groups are overlaid with the corresponding empirical isochrones obtained from SLFR analysis (Secs.~\ref{sec:Isochrones}, \ref{subsec:ageanalysis}). In the right panel, the data are overlaid with theoretical isochrones from \citet{Tognelli2018} for ages of 5, 10, and 20 Myr. }
    \label{fig:compiso}
\end{figure}

Like other nearby stellar groups, the presently known membership of $\epsilon$CA is dominated by M stars, and the spectral type distribution peaks in the mid-M range. The multiplicity fraction of $\epsilon$CA (40\%) is intermediate between those of the field star population and T associations. We find an overall circumstellar disk fraction of 30\% for $\epsilon$CA, with the vast majority of disk-bearing stars lying within $\sim$5 pc (Figures~\ref{fig:finaluvw} and \ref{fig:diskdist}). 

In providing a well-defined region of $XYZ$ and $UVW$ space encompassing this NYMG, this work sets the stage for a more complete search of Gaia DR2 (and eventual DR3) data for new members of $\epsilon$CA. This work thereby provides the framework for future investigations of the initial mass function as well as multiplicity and disk fractions of $\epsilon$CA. 

\acknowledgements

We thank the anonymous referee for numerous comments that improved this manuscript. We acknowledge helpful comments from Jonathan Gagn\'{e} and Eric Mamajek. This research is supported by NASA Exoplanets Program grant 80NSSC19K0292 to Rochester Institute of Technology. This work has made use of data from the European Space Agency (ESA) mission {\it Gaia} (\url{https://www.cosmos.esa.int/gaia}), processed by the {\it Gaia} Data Processing and Analysis Consortium (DPAC, \url{https://www.cosmos.esa.int/web/gaia/dpac/consortium}). Funding for the DPAC has been provided by national institutions, in particular the institutions participating in the {\it Gaia} Multilateral Agreement. This publication makes use of data products from the Two Micron All Sky Survey, which is a joint project of the University of Massachusetts and the Infrared Processing and Analysis Center/California Institute of Technology, funded by the National Aeronautics and Space Administration and the National Science Foundation.

\bibliography{main}
\bibliographystyle{aasjournal}

\begin{longrotatetable}
\begin{deluxetable*}{lrrrrrrrrrrrrrrrr}
\tablecaption{Gaia Data and RVs for Candidate $\epsilon$CA Members}
\tabletypesize{\scriptsize}
\label{table:BOT}
\tablehead{\colhead{Name$^a$} & \colhead{RA (J2015.5)} & \colhead{Dec (J2015.5)} & \colhead{$\pi$} & \colhead{$\sigma$} &\colhead{G} & \colhead{$\sigma$} & \colhead{G$-$R$_{P}$} & \colhead{$\sigma$} & \colhead{PMRA} & \colhead{$\sigma$} & \colhead{PMDec} & \colhead{$\sigma$} & \colhead{RV} & \colhead{$\sigma$}& \colhead{RUWE$^b$} & \colhead{RV ref$^c$}\\ \colhead{} & \colhead{(deg)} & \colhead{(deg)} & \colhead{(mas)} & \colhead{} & \colhead{(mag)} & \colhead{} & \colhead{(mag)} & \colhead{} & \colhead{(mas/yr)} & \colhead{} & \colhead{(mas/yr)} & \colhead{} & \colhead{(km/s)} & \colhead{}}
\startdata
HD 8279$*$ & 142.0874 & -78.2597 & 8.504 & 0.025 & 8.92760 & 0.00038 & 0.3737 & 0.0025 & -28.211 & 0.043 & 20.350 & 0.046 & 12.76 & 0.2 & 1.05 & 2\\ 
CP-68 1388 $\dagger$ & 164.4553 & -69.2333 & 8.592 & 0.029 & 10.0229 & 0.0024 & 0.6763 & 0.0070 & -34.842 & 0.051 & 3.552 & 0.044 & 15.9 & 1.0 & 0.93 & 12 \\
VW Cha$*$ & 167.0058 & -77.7079 & \nodata & \nodata & 12.187 & 0.020 & 1.425 & 0.018 & \nodata & \nodata & \nodata & \nodata & \nodata & \nodata & \nodata \\
TYC 9414-191-1$\ddagger$ & 169.1200 & -78.4224 & 1.564 & 0.022 & 10.523674 & 0.00025 &  0.86718 & 0.00067 & -38.372 & 0.040 & 2.460 & 0.033 & \nodata & \nodata & 0.91 & \nodata\\
2MASS J11183572-7935548 $\dagger$ & 169.6478 & -79.5986 & 10.57 & 0.15 & 13.8687 & 0.0027 & 1.3726 & 0.0043 & -41.76 & 0.26 & 4.98 & 0.24 & 19.3 & 1.6 & 3.35$^+$ & 8 \\ 
RX J1123.2-7924$*$ & 170.7311 & -79.4123 & 9.379 & 0.021 & 12.8674 & 0.0016 & 1.0036 & 0.0041 & -31.670 & 0.040 & -17.443 & 0.038 & 2.7 & 2.9 & 1.14 & 8 \\ 
HIP 55746$*$  & 171.3217 & -84.9545 & 11.077 & 0.138 & 7.53435 & 0.00044 & 0.3524 & 0.0021 & -46.19 & 0.16 & 11.92 & 0.16 & 20.9 & 1.2 & 1.46 & 3\\ 
2MASS J11334926-7618399$*$ & 173.4550 & -76.3111 & 5.410 & 0.060 & 15.56704 & 0.00081 & 1.3373 & 0.0010 & -22.41 & 0.11 & -0.561 & 0.087 & \nodata & \nodata & 1.07 & \nodata \\
RX J1137.4-7648$*$  & 174.3794 & -76.7997 & 11.988 & 0.021 & 13.52653 & 0.00079 & 1.0494 & 0.0022 & -60.544 & 0.041 & -8.935 & 0.032 & 14.0 & 5.0 & 1.16 & 8 \\ 
2MASS J11404967-7459394 $\dagger$ & 175.2063 & -74.9942 & 10.397 & 0.069 & 16.4291 & 0.0018 & 1.4466 & 0.0044 & -42.74 & 0.13 & -2.28 & 0.10 & 10.3 & 1.0 & 0.94 & 8 \\ 
2MASS J11411722-7315369\# & 175.3211 & -73.2602 & 10.416 & 0.052 & 15.47804 & 0.00084 & 1.3229 & 0.0024 & -42.047 & 0.093 & 5.672 & 0.082 & \nodata & \nodata & 1.07 & \nodata\\ 
TYC 9238-612-1$*$ & 175.3646 & -73.7841 & 6.124 & 0.021 & 10.5976 & 0.0012 & 0.5663 & 0.0035 & -35.747 & 0.041 & 10.938 & 0.036 & \nodata & \nodata & 1.33 & \nodata\\
2MASS J11432669-7804454 $\dagger$ & 175.8603 & -78.0793 & 5.54 & 0.56 & 14.8931 & 0.0072 & 1.321 & 0.022 & -43.98 & 0.92 & -6.52 & 0.75 & 15.6 & 1 & 10.51$^+$ & 8 \\
2MASS J11432968-7418377\# & 175.8729 & -74.3105 & 10.063 & 0.039 & 14.17633 & 0.00066 & 1.2213 & 0.0022 & -40.621 & 0.073 & -3.026 & 0.067 & \nodata & \nodata & 1.19 & \nodata\\ 
RX J1147.7-7842 $\dagger$ & 176.9496 & -78.698 & 9.894 & 0.039 & 12.3523 & 0.0012 & 1.1840 & 0.0028 & -41.660 & 0.075 & -4.265 & 0.065 & 16.1 & 0.9 & 1.28 & 8\\ 
RX J1149.8-7850 $\dagger$ & 177.3818 & -78.8503 & 9.918 & 0.025 & 11.9324 & 0.0043 & 1.028 & 0.015 & -41.876 & 0.047 & -4.265 & 0.040 & 13.4 & 1.3 & 1.19  & 8\\ 
RX J1150.4-7704 $*$ & 177.6171 & -77.0773 & 6.551 & 0.031 & 11.5250 & 0.0016 & 0.7773 & 0.0042 & -42.744 & 0.054 & -11.047 & 0.040 & \nodata &\nodata & 1.26 & \nodata \\
RX J1150.9-7411 $\dagger$ & 177.6868 & -74.1871 & 10.63 & 0.18 & 13.4567 & 0.0050 & 1.324 & 0.012 & -39.50 & 0.32 & -8.08 & 0.29 & 15.0 & 1.2 & 6.01$^+$ & 8\\ 
2MASS J11550485-7919108 $\dagger$ & 178.7692 & -79.3198 & 9.886 & 0.058 & 14.8180 & 0.0017 & 1.3502 & 0.0053 & -41.18 & 0.13 & -4.336 & 0.086 & \nodata & \nodata & 1.08 & \nodata\\ 
T Cha $\dagger$ & 179.3054 & -79.3588 & 9.122 & 0.083 & 12.97 & 0.11 & 1.61 & 0.35 & -42.00 & 0.12 & -9.245 & 0.080 & 6--30 & \nodata & 0.91 & 2,5,10 \\ 
RX J1158.5-7754B $\dagger$ & 179.6109 & -77.9126 & 9.662 & 0.035 & 13.2023 & 0.0027 & 1.1713 & 0.0081 & -39.573 & 0.064 & -5.686 & 0.060 & 13.0 & 2.0 & 1.22 & 11 \\ 
RX J1158.5-7754A $\dagger$ & 179.6165 & -77.9083 & 9.518 & 0.035 & 9.9798 & 0.0011 & 0.7888 & 0.0033 & -39.660 & 0.063 & -12.844 & 0.070 & 14.02 & 1.82 & 1.66 & 2,8,12\\ 
HD 104036 $\dagger$ & 179.646 & -77.8254 & 9.566 & 0.038 & 6.69296 & 0.00053 & 0.1646 & 0.0041 & -41.286 & 0.065 & -7.762 & 0.073 & 12.6 & 0.5 & 1.23 & 3\\ 
CXOU J115908.2-781232$\ddagger$ & 179.7824 & -78.2089 & 9.425 & 0.062 & 15.2563 & 0.00063 & 1.3139 & 0.0022 & -38.62 & 0.14 & -5.181 & 0.091 & 15.1 & 0.2 & 1.24 & 8\\ 
$\epsilon$ Cha AB$\dagger$ & 179.9056 & -78.2219 & \nodata & \nodata & 4.7816 & 0.0080 & -0.1505 & 0.0086 & \nodata & \nodata & \nodata & \nodata & 13 & 3.7 & \nodata & 11\\
RX J1159.7-7601 $\dagger$ & 179.9254 & -76.024 & 10.025 & 0.023 & 10.8095 & 0.0023 & 0.7694 & 0.0058 & -41.025 & 0.043 & -6.190 & 0.038 & 13.0 & 3.7 & 0.96 & 2,8\\ 
2MASS J12000269-7444068\# & 180.0105 & -74.7352 & 10.067 & 0.049 & 14.3241 & 0.0011 & 1.2431 & 0.0028 & -42.047 & 0.079 & -5.699 & 0.071 & \nodata & \nodata & 1.24 & \nodata\\ 
HD 104237A $\dagger$ & 180.0204 & -78.193 & 9.226 & 0.058 & 6.5427 & 0.0018 & 0.2213 & 0.0074 & -39.31 & 0.11 & -6.212 & 0.083 & 13.52 & 0.39 & 1.55 & 4 \\ 
HD 104237D $\dagger$ & 180.0336 & -78.1943 & 9.885 & 0.061 & 13.0206 & 0.0015 & 1.1981 & 0.0065 & -38.87 & 0.13 & -3.195 & 0.098 & \nodata & \nodata & 1.53 & \nodata\\ 
HD 104237E $\dagger$ & 180.0379 & -78.1951 & 9.796 & 0.034 & 11.838 & 0.028 & 1.16 & 0.11 & -42.907 & 0.074 & -4.418 & 0.051 & \nodata & \nodata & 1.54 & \nodata \\ 
2MASS J12005517-7820296 $\dagger$ & 180.2291 & -78.3415 & 9.713 & 0.082 & 15.5826 & 0.0012 & 1.4005 & 0.0040 & -40.59 & 0.14 & -4.96 & 0.12 & 10.7 & 1.3 & 1.17 & 8\\ 
HD 104467 $\dagger$ & 180.412 & -78.9881 & 10.14 & 0.17 & 8.4285 & 0.0022 & 0.5145 & 0.0059 & -41.11 & 0.26 & -5.41 & 0.26 & 12.81 & 0.96 & 5.17 & 2 \\ 
2MASS J12014343-7835472 $\dagger$ & 180.4302 & -78.5965 & 9.53 & 0.11 & 17.115 & 0.021 & 1.166 & 0.069 & -41.32 & 0.19 & -6.33 & 0.14 & 20.0 & 0.6 & 1.28 & 8\\ 
USNO-B 120144.7-781926 $\dagger$ & 180.4343 & -78.3241 & 9.819 & 0.063 & 15.2758 & 0.0076 & 1.372 & 0.020 & -41.43 & 0.11 & -6.105 & 0.088 & 14.9 & 1.1 & 1.08 & 8\\ 
CXOU J120152.8-781840 $\dagger$ & 180.4681 & -78.3115 & 9.735 & 0.068 & 14.9821 & 0.0012 & 1.3319 & 0.0026 & -40.59 & 0.12 & -6.88 & 0.10 & 16.5 & 1.1 & 1.14 & 8\\ 
RX J1202.1-7853 $\dagger$ & 180.5145 & -78.8837 & 10.011 & 0.046 & 11.5100 & 0.0030 & 1.0059 & 0.0088 & -45.122 & 0.081 & -4.124 & 0.069 & 17.1 & 0.2 & 1.63 & 5\\ 
RX J1202.8-7718$\ddagger$ & 180.7269 & -77.3106 & 9.606 & 0.035 & 13.3749 & 0.0026 & 1.1762 & 0.0080 & -39.625 & 0.072 & -6.044 & 0.059 & 14.4 & 0.6 & 1.21 & 7\\ 
RX J1204.6-7731 $\dagger$ & 181.1498 & -77.5263 & 9.922 & 0.034 & 12.5714 & 0.0018 & 1.1513 & 0.0053 & -41.364 & 0.056 & -6.397 & 0.052 & 10.4 & 2.0 & 1.20 & 6 \\ 
TYC 9420-676-1 $*$ & 181.2384 & -79.5346 & 3.700 & 0.080 & 10.23947 & 0.00063 & 0.4443 & 0.0016 & -38.16 & 0.16 & -1.00 & 0.18 & \nodata & \nodata & 2.95 & \nodata\\
HD 105234$\ddagger$ & 181.7721 & -78.7412 & 9.566 & 0.035 & 7.41947 & 0.00041 & 0.2247 & 0.0028 & -40.971 & 0.058 & -9.41 & 0.057 & \nodata & \nodata & 1.15 & \nodata\\ 
2MASS J12074597-7816064$*$ & 181.9407 & -78.2685 & 9.454 & 0.058 & 14.55144 & 0.00073 & 1.2293 & 0.0021 & -38.424 & 0.091 & -6.319 & 0.077 & 15.2 & 1.9 & 1.33 & 8\\ 
RX J1207.7-7953 $\dagger$ & 181.9501 & -79.8785 & 10.016 & 0.030 & 13.3403 & 0.0022 & 1.1860 & 0.0057 & -42.058 & 0.056 & -7.113 & 0.052 & 15.0 & 0.7 & 1.12 & 8\\
HIP 59243$\ddagger$ & 182.2811 & -78.7813 & 10.019 & 0.033 & 6.79373 & 0.00043 & 0.1889 & 0.0034 & -43.364 & 0.062 & -7.619 & 0.06 & \nodata & \nodata & 1.10 & \nodata\\ 
HD 105923 $\dagger$ & 182.9084 & -71.1767 & 9.400 & 0.036 & 8.83653 & 0.00083 & 0.5367 & 0.0035 & -38.721 & 0.062 & -7.423 & 0.051 & 14.34 & 1.06 & 0.87 & 1,2,12\\ 
RX J1216.8-7753 $\dagger$ & 184.1906 & -77.8927 & 9.820 & 0.039 & 12.9206 & 0.0037 & 1.132 & 0.010 & -39.825 & 0.072 & -9.074 & 0.072 & 14.0 & 2.0 & 1.17 & 6\\ 
RX J1219.7-7403 $\dagger$ & 184.9315 & -74.0659 & 9.865 & 0.027 & 12.1400 & 0.0016 & 1.0187 & 0.0042 & -40.323 & 0.043 & -9.256 & 0.039 & 13.86 & 1.89 & 0.96 & 2 \\ 
RX J1220.4-7407 $\dagger$ & 185.0901 & -74.1277 & 6.71 & 0.70 & 11.9406 & 0.0020 & 1.0706 & 0.0050 & -40.9 & 1.2 & -4.1 & 1.4 & 12.3 & 0.4 & 38.28$^+$ & 6\\
2MASS J12203396-7135188\# & 185.1411 & -71.5886 & 10.773 & 0.057 & 13.21947 & 0.00084 & 1.2486 & 0.0019 & -42.573 & 0.096 & -8.571 & 0.070 & \nodata & \nodata & 1.28 & \nodata\\ 
2MASS J12203619-7353027\# & 185.1502 & -73.8842 & 10.007 & 0.035 & 13.1409 & 0.0010 & 1.1424 & 0.0024 & -40.621 & 0.069 & -9.326 & 0.050 & \nodata & \nodata & 1.12 & \nodata\\ 
2MASS J12210499-7116493 $\dagger$ & 185.2703 & -71.2804 & 10.055 & 0.024 & 11.2032 & 0.0028 & 0.8874 & 0.0079 & -40.416 & 0.043 & -9.647 & 0.034 & 11.44 & 2.53 & 1.02 & 9 \\ 
2MASS J12220068-7001041\# & 185.5022 & -70.0178 & 10.008 & 0.047 & 14.0236 & 0.0011 & 1.2371 & 0.0045 & -41.229 & 0.096 & -10.474 & 0.063 & \nodata & \nodata & 1.23 & \nodata\\ 
2MASS J12222238-7137040\# & 185.5926 & -71.6178 & 10.516 & 0.027 & 12.5299 & 0.0081 & 1.105 & 0.023 & -41.843 & 0.044 & -9.962 & 0.043 & \nodata & \nodata & 1.06 & \nodata\\ 
2MASS J12224862-7410203\# & 185.702 & -74.1724 & 10.605 & 0.089 & 16.1751 & 0.0024 & 1.4774 & 0.0068 & -42.57 & 0.15 & -9.27 & 0.13 & \nodata & \nodata & 1.18 & \nodata\\ 
2MASS J12255824-7551116\# & 186.4921 & -75.8533 & 10.049 & 0.065 & 15.36554 & 0.00074 & 1.3318 & 0.0026 & -41.23 & 0.11 & -10.46 & 0.10 & \nodata & \nodata & 1.20 & \nodata\\
2MASS J12324805-7654237\# & 188.1995 & -76.9067 & 11.69 & 0.91 & 12.8346 & 0.0012 & 1.1960 & 0.0033 & -41.84 & 1.67 & -13.49 & 1.32 & \nodata & \nodata & 36.83$^+$ & \nodata\\
2MASS J12332483-6848553\# & 188.3531 & -68.8154 & 10.061 & 0.086 & 15.7118 & 0.0012 & 1.3676 & 0.0039 & -41.43 & 0.13 & -12.51 & 0.11 & \nodata & \nodata & 1.16 & \nodata\\ 
2MASS J12351540-7043079\# & 188.8135 & -70.7189 & 10.705 & 0.066 & 15.70148 & 0.00068 & 1.3844 & 0.0026 & -43.72 & 0.11 & -12.639 & 0.094 & \nodata & \nodata & 1.03 & \nodata\\ 
RX J1239.4-7502 $\dagger$ & 189.838 & -75.0443 & 9.646 & 0.027 & 9.9751 & 0.0017 & 0.6679 & 0.0048 & -38.214 & 0.044 & -12.368 & 0.042 & 13.62 & 2.8 & 0.97 & 9\\ 
2MASS J12421315-6943484\# & 190.5543 & -69.7302 & 10.062 & 0.083 & 15.1444 & 0.0011 & 1.3551 & 0.0039 & -39.50 & 0.14 & -12.77 & 0.10 & \nodata & \nodata & 1.24 & \nodata\\ 
RX J1243.1-7458$*$  & 190.7219 & -74.98 & 7.99 & 0.37 & 13.7652 & 0.0044 & 1.186 & 0.011 & -17.25 & 0.66 & -2.63 & 0.53 & 13.5 & 7.0 & 10.71$^+$ & 8\\ 
2MASS J12425584-7034207\# & 190.7321 & -70.5725 & 10.062 & 0.078 & 15.97496 & 0.00098 & 1.4049 & 0.0031 & -40.47 & 0.12 & -13.20 & 0.12 & \nodata & \nodata & 0.99 & \nodata\\ 
2MASS J12473611-7031135\# & 191.9000 & -70.5206 & 10.012 & 0.031 & 13.26920 & 0.00065 & 1.1498 & 0.0016 & -40.249 & 0.048 & -14.241 & 0.045 & \nodata & \nodata & 1.12 & \nodata\\
CD-69 1055 $\dagger$ & 194.6061 & -70.4804 & 10.549 & 0.029 & 9.6062 & 0.0027 & 0.6388 & 0.0076 & -41.001 & 0.047 & -16.459 & 0.045 & 11.18 & 1.67 & 1.01 & 1,2,8\\ 
CM Cha$\ddagger$ & 195.5560 & -76.6328 & 5.152 & 0.024 & 12.5599 & 0.0043 & 1.001 & 0.012 & -20.990 & 0.040 & -9.144 & 0.041 & \nodata & \nodata & 1.15  & \nodata \\
MP Mus $\dagger$ & 200.531 & -69.6368 & 10.115 & 0.031 & 9.9521 & 0.0030 & 0.7207 & 0.0087 & -38.289 & 0.044 & -20.204 & 0.045 & 11.6 & 0.2 & 0.98 & 12\\ 
\enddata
\tablecomments{a) M+13 \textit{bona fide} members are indicated by $\dagger$, M+13 provisional members are indicated by $\ddagger$, M+13 rejected members are indicated by $*$, and GF18 members are indicated by \#.\\ b) Values marked with $+$ have high astrometric excess noise.\\ c) References for radial velocities: $^1$\citet{Desidera2015} $^2$\citet{Gaia2018} $^3$\citet{Gontcharov2006}, $^4$\citet{Grady2004}, $^5$\citet{Guenther2007}, $^6$\citet{Lopez2013}, $^7$\citet{Malo2014}, $^8$\citet{Murphy2013}, $^9$\citet{RAVE2013}, $^{10}$\citet{Schisano2009},$^{11}$\citet{Terranegra1999}, $^{12}$\citet{Torres2006}}
\end{deluxetable*}
\end{longrotatetable}

\begin{deluxetable*}{lrrrrrrr}
\tablecaption{Membership Quality Flags}
\tabletypesize{\scriptsize}
\label{table:deltas}
\tablehead{\colhead{Name} & \colhead{$K_{kin}$} & \colhead{$\sigma_{K_{kin}}$} & \colhead{$\Delta M$} &\colhead{$\sigma_{\Delta M}$} &\colhead{Status} & \colhead{Notes}}
\startdata
HD 8279 & 4.35 & 0.21 & 1.8911 & 0.0066 & R & Previously rejected (Sec~\ref{subsec:outliers})\\
CP-68 1388 & 2.0 & 1.0 & -0.1975 & 0.0097 & M \\
2MASS J11183572-7935548A & 4.7 & 1.6 & -1.200 & 0.033 & M\\
RX J1123.2-7924 & 15.8 & 2.9 & 0.9509 & 0.0065 & R & Previously rejected (Sec~\ref{subsec:outliers})\\
HIP 55746 & 7.1 & 1.2 & 1.383 & 0.027 & R & Previously rejected (Sec~\ref{subsec:outliers})\\
RX J1137.4-7648 & 4.6 & 5.0 & 1.8790 & 0.0046 & R & Previously rejected (Sec~\ref{subsec:outliers})\\
2MASS J11404967-7459394 & 4.2 & 1.0 & 0.025 & 0.016 & M\\
2MASS J11411722-7315369 & \nodata & \nodata & 1.093 & 0.012 & P \\
2MASS J11432669-7804454 & 17.9 & 4.1 & -0.83 & 0.24 & R & Probable Cha Cloud (Sec~\ref{subsec:outliers})\\
2MASS J11432968-7418377 & \nodata & \nodata & 0.8659 & 0.0090 & P \\
RX J1147.7-7842 & 1.81 & 0.90 & -0.6586 & 0.0097 & M \\
RX J1149.8-7850 & 1.0 & 1.3 & -0.0002 & 0.0097 & M\\
RX J1150.9-7411 & 2.7 & 1.3 & -0.892 & 0.042 & M\\
2MASS J11550485-7919108 & \nodata& \nodata& -0.059 & 0.015 & M \\
T Cha & 3.0 & 1.3 & -7.72 & 0.13 & M & Anomalous CMD position (Sec~\ref{sec:retained})\\
RX J1158.5-7754B & 1.3 & 2.0 & 0.2454 & 0.0105 & M\\
RX J1158.5-7754A & 3.6 & 1.8 & -0.7542 & 0.0091 & M\\
HD 104036 & 1.87 & 0.51 & 3.5594 & 0.0092 & M\\
CXOU J115908.2-781232 & 1.22 & 0.25 & 0.774 & 0.015 & M \\
RX J1159.7-7601 & 1.3 & 3.7 & 0.3029 & 0.0073 & M\\
2MASS J12000269-7444068 & \nodata & \nodata & 0.799 & 0.012 & P \\
HD 104237A & 0.69 & 0.42 & 2.21 & 0.016 & M\\
HD 104237D & \nodata& \nodata& -0.115 & 0.015 & M\\
HD 104237E & \nodata& \nodata& -1.027 & 0.036 & M\\
2MASS J12005517-7820296 & 3.4 & 1.3 & -0.125 & 0.020 & M\\
HD 104467 & 1.6 & 1.0 & 0.010 & 0.038 & M\\
2MASS J12014343-7835472 & 5.93 & 0.65 & 4.168 & 0.045 & M & Anomalous CMD position (Sec~\ref{sec:retained})\\
USNO-B 120144.7-781926 & 0.83 & 1.11 & 0.064 & 0.022 & M\\
CXOU J120152.8-781840 & 2.5 & 1.1 & 0.33 & 0.0163 & M\\
RX J1202.1-7853 & 3.35 & 0.23 & -0.279 & 0.013 & M\\
RX J1202.8-7718 & 0.77 & 0.61 & 0.365 & 0.010 & M \\
RX J1204.6-7731 & 3.6 & 2.0 & -0.1695 & 0.0091 & M\\
HD 105234 & \nodata & \nodata & 3.1047 & 0.0083 & P \\
2MASS J12074597-7816064 & 1.5 & 1.9 & 1.029 & 0.014 & R & Previously rejected (Sec~\ref{subsec:outliers})\\
RX J1207.7-7953 & 1.01 & 0.70 & 0.3388 & 0.0087 & M\\
HIP 59243 & \nodata & \nodata & 3.2697 & 0.0075 & P \\
HD 105923 & 0.9 & 2.0 & 0.0231 & 0.0091 & M\\
RX J1216.8-7753 & 1.3 & 1.9 & 0.304 & 0.012 & M\\
RX J1219.7-7403 & 9.2 & 3.3 & 0.2485 & 0.0075 & M\\
RX J1220.4-7407 & 2.1 & 1.4 & -1.09 & 0.24 & R & Probable Cha Cloud (Sec~\ref{subsec:outliers})\\
2MASS J12203396-7135188 & \nodata & \nodata & -0.216 & 0.012 & P \\
2MASS J12203619-7353027 & \nodata & \nodata & 0.4854 & 0.0086 & P \\
2MASS J12210499-7116493 & 2.6 & 2.5 & 0.0482 & 0.0080 & M \\
2MASS J12220068-7001041 & \nodata & \nodata & 0.547 & 0.011 & P \\
2MASS J12222238-7137040 & \nodata & \nodata & 0.248 & 0.014 & P \\
2MASS J12224862-7410203 & \nodata & \nodata & -0.822 & 0.021 & P \\
2MASS J12255824-7551116 & \nodata & \nodata & 0.784 & 0.015 & P \\
2MASS J12324805-7654237 & \nodata & \nodata & 0.08 & 0.18 & P \\
2MASS J12332483-6848553 & \nodata & \nodata & 0.613 & 0.020 & P\\
2MASS J12351540-7043079 & \nodata & \nodata & 0.473 & 0.014 & P\\
RX J1239.4-7502 & 1.5 & 2.8 & 0.0672 & 0.0077 & M\\
2MASS J12421315-6943484 & \nodata & \nodata & 0.235 & 0.019 & P \\
RX J1243.1-7458 & 11.1 & 7.0 & 0.27 & 0.11 & R & Previously rejected (Sec~\ref{subsec:outliers})\\
2MASS J12425584-7034207 & \nodata & \nodata & 0.270 & 0.018 & P \\
2MASS J12473611-7031135 & \nodata & \nodata & 0.5594 & 0.0073 & P \\
CD-69 1055 & 2.5 & 1.9 & 0.1176 & 0.0087 & M\\
MP Mus & 2.32 & 0.21 & -0.2265 & 0.0097 & M\\
\enddata
\tablecomments{Key for status column: M = member; P = provisional member; R = rejected.}
\end{deluxetable*}

\begin{deluxetable*}{lrrrrrrrrrrr}
\tablecaption{Final $\epsilon$ Cha Association Membership List$^a$}
\tabletypesize{\scriptsize}
\label{table:finalmemb}
\tablehead{\colhead{Name} & \colhead{SpT} & \colhead{Distance [pc]$^b$} & \colhead{G$_{BP}$} &\colhead{G} & \colhead{G$_{RP}$} & \colhead{J} & \colhead{H} & \colhead{K} & \colhead{Multiplicity$^c$} & \colhead{Disk?$^d$}}
\startdata
CP-68 1388 & K1 & $115.66_{-0.38}^{+0.39}$ & 10.568 & 10.023 & 9.347 & 8.48 & 8.01 & 7.79 & \nodata & \nodata\\
2MASS J11183572-7935548A & M4.5 & $94.14_{-1.30}^{+1.34}$ & 15.513 & 13.869 & 12.50 & 10.50 & 9.89 & 9.62 & P & \nodata\\
2MASS J11183572-7935548B* & \nodata & $93.04_{-3.35}^{+3.61}$ & \nodata & 16.189 & \nodata & \nodata & \nodata & \nodata &\nodata & \nodata\\
2MASS J11404967-7459394 & M5.5 & 95.68$_{-0.64}^{+0.64}$ & 18.833 & 16.429 & 14.983 & 12.68 & 12.15 & 11.77 & \nodata & \nodata\\
RX J1147.7-7842 & M3.5 & $100.52_{-0.39}^{+0.39}$ & 13.814 & 12.352 & 11.168 & 9.52 & 8.86 & 8.59 & P & \nodata\\
RX J1149.8-7850 & M0 & $100.28_{-0.25}^{+0.25}$ & 13.0419 & 11.932 & 10.904 & 9.45 & 8.72 & 8.49 & \nodata & Y \\
RX 1150.9-7411 & M3.7 & $93.61_{-1.56}^{+1.62}$ & 14.852 & 13.457 & 12.133 & 10.38 & 9.78 & 9.48 & VP & \nodata\\
2MASS J11550485-7919108 & M3 & $100.60_{-0.59}^{+0.60}$ & 16.826 & 14.818 & 13.468 & 11.22 & 10.47 & 10.08 & C & Y\\
T Cha & G8 & $108.98_{-0.97}^{+0.99}$ & 13.682 & 12.974 & 11.368 & 8.96 & 7.86 & 6.95 & \nodata & Y\\
RX J1158.5-7754B & M3 & $102.92_{-0.37}^{+0.37}$ & 14.629 & 13.202 & 12.031 & 10.34 & 9.72 & 9.44 & \nodata & \nodata\\
RX J1158.5-7754A & K4 & $104.47_{-0.38}^{+0.39}$ & 10.688 & 9.980 & 9.191 & 8.22 & 7.56 & 7.40 & VP & \nodata\\
HD 104036 & A7 & $103.95_{-0.41}^{+0.42}$ & 6.806 & 6.693 & 6.528 & 6.29 & 6.22 & 6.11 & V & \nodata\\
CXOU  J115908.2-781232* & M4.75 & $105.50_{-0.68}^{+0.69}$ & 17.118 & 15.256 & 13.942 & 12.01 & 11.45 & 11.17 & \nodata & Y\\
$\epsilon$ Cha AaAbB & B9 & 111H & 4.862 & 4.782 & 4.931 & 5.02 & 5.04 & 4.98 & V & \nodata\\
RX J1159.7-7601 & K4 & $99.21_{-0.23}^{+0.23}$ & 11.491 & 10.809 & 10.040 & 9.14 & 8.47 & 8.30 & \nodata & \nodata\\
HD 104237C & M/L & \nodata & \nodata & \nodata & \nodata & \nodata & \nodata & \nodata & \nodata & \nodata\\
HD 104237B & K/M & \nodata & \nodata & \nodata & \nodata & \nodata & \nodata & \nodata & \nodata & \nodata\\ 
HD 104237A & A7.75 & $107.76_{-0.67}^{+0.68}$ & 6.710 & 6.543 & 6.321 & 5.81 & 5.25 & 4.59 & \nodata & Y\\
HD 104237D & M3.5 & $100.61_{-0.62}^{+0.62}$ & 14.440 & 13.021 & 11.823 & 9.62 & 8.74 & 8.12 & \nodata & ?\\
HD 104237E & K5.5 & $101.52_{-0.34}^{+0.35}$ & 12.653 & 11.838 & 10.675 & $\ge$9.10 & $\ge$8.25 & 7.49 & P & Y\\
2MASS J12005517-7820296 & M5.75 & $102.39_{-0.85}^{+0.87}$ & 17.865 & 15.583 & 14.182 & 11.96 & 11.40 & 11.01 & S & Y\\
HD 104467 & G3 & $98.11_{-1.56}^{+1.62}$ & 8.796 & 8.428 & 7.914 & 7.26 & 6.97 & 6.85 & C & \nodata\\
2MASS J12014343-7835472 & M2.25 & $104.334_{-1.13}^{+1.16}$	& 18.205 & 17.115 & 15.949 & 14.36 & 13.38 & 12.81 & \nodata & Y\\
USNO-B 120144.7-781926 & M5 & $101.29_{-0.64}^{+0.65}$ & 17.248 & 15.276 & 13.904 & 11.68 & 11.12 & 10.78 & \nodata & Y\\
CXOU J120152.8-781840 & M4.75 & $102.16_{-0.70}^{+0.71}$ & 16.939 & 14.982 & 13.650 & 11.63 & 11.04 & 10.77 & \nodata & \nodata\\
RX J1202.1-7853 & M0 & $99.36_{-0.45}^{+0.46}$ & 12.547 & 11.510 & 10.504 & 9.215 & 8.46 & 8.31 & V & \nodata \\
RX  J1202.8-7718* & M3.5 & $103.52_{-0.37}^{+0.37}$ & 14.795 & 13.375 & 12.199 & 10.51 & 9.83 & 9.59 & \nodata & \nodata\\
RX J1204.6-7731 & M3 & $100.24_{-0.33}^{+0.34}$ & 13.94 & 12.571 & 11.420 & 9.77 & 9.13 & 8.88 & \nodata & \nodata\\
RX J1207.7-7953 & M3.5 & $99.30_{-0.30}^{+0.30}$ & 14.793 & 13.340 & 12.154 & 10.43 & 9.76 & 9.57 & \nodata & \nodata\\
HD 105923 & G8 & $105.78_{-0.40}^{+0.41}$ & 9.245 & 8.837 & 8.300 & 7.67 & 7.31 & 7.18 & VC & \nodata\\
RX J1216.8-7753 & M4 & $101.27_{-0.40}^{+0.40}$ & 14.236 & 12.921 & 11.789 & 10.09 & 9.47 & 9.24 & \nodata & \nodata\\
RX J1219.7-7403 & M0 & $100.82_{-0.27}^{+0.27}$	& 13.201 & 12.140 & 11.121 & 9.75 & 9.05 & 8.86 & \nodata & \nodata\\
2MASS J12210499-7116493 & K7 & $98.92_{-0.23}^{+0.24}$ & 12.044 & 11.203 &10.316 & 9.09 & 8.42 & 8.24 & \nodata & \nodata\\
RX J1239.4-7502 & K3 & $103.10_{-0.28}^{+0.29}$ & 10.531 & 9.975 & 9.307 & 8.43 & 7.95 & 7.78 & \nodata & \nodata\\
CD-69 1055 & K0 & $94.32_{-0.26}^{+0.26}$ & 10.132 & 9.606 & 8.967 & 8.18 & 7.70 & 7.55 & \nodata & \nodata\\
MP Mus & K1 & $98.34_{-0.30}^{+0.30}$ & 10.547 & 9.952 & 9.231 & 8.28 & 7.64 & 7.29 & \nodata & Y\\
\enddata
\tablecomments{a) * = previously not considered a \textit{bona fide} member (M+13). \\ b) Distances have been calculated using the inverse parallax method with zero-point corrections \citep{Lindegren2018}. \\ c) S = suspected spectroscopic binary (M+13); V = resolved binary \citep{Briceno2017}; P = possible photometric binary (this work); C = potential wide separation companion (this work). \\ d) IR excess and/or other evidence indicative of presence of circumstellar disk (M+13 and references therein).}
\end{deluxetable*}

\begin{deluxetable*}{lrrrrrrrrrrr}
\tablecaption{Provisional $\epsilon$ Cha Association Membership List}
\tabletypesize{\scriptsize}
\label{table:provmemb}
\tablehead{\colhead{Name} & \colhead{SpT$^a$} & \colhead{Distance [pc]$^b$} & \colhead{G$_{BP}$} &\colhead{G} & \colhead{G$_{RP}$} & \colhead{J} & \colhead{H} & \colhead{K} & \colhead{Multiplicity$^c$} & \colhead{Disk?$^d$}}
\startdata
2MASS J11411722-7315369 & M4.5 & 95.51$_{-0.47}^{+0.47}$ & 17.382 & 15.478 & 14.155 & 12.16 & 11.56 & 11.31 & \nodata & \nodata\\
2MASS J11432968-7418377 & M3.5 & 98.84$_{-0.38}^{+0.38}$ & 15.742 & 14.176 & 12.955 & 11.25 & 10.60 & 10.37 & \nodata & \nodata\\
2MASS J11550336-7919147 & M6 & $101.26_{-5.12}^{+5.70}$ & 20.476 & 19.925 & 18.189 & 15.85 & 15.03 & $\ge$12.64 & \nodata & Y \\
2MASS J12000269-7444068 & M4 & 98.80$_{+0.48}^{-0.48}$ & 15.959 & 14.324 & 13.081 & 11.37 & 10.67 & 10.42 & \nodata & \nodata\\
2MASS J12011981-7859057 & M5 & $101.55_{-0.77}^{+0.78}$ & 17.251 & 15.179 & 13.823 & 11.75 & 11.20 & 10.89 & \nodata & \nodata\\
HD 105234 & A9$\ddagger$ & $103.96_{-0.37}^{+0.38}$ & 7.570 & 7.419 & 7.195 & 6.87 & 6.76 & 6.68 & V & Y\\
HIP 59243 & A6$\ddagger$ & $99.27_{-0.32}^{+0.32}$ & 6.934 & 6.794 & 6.605 & 6.35 & 6.23 & 6.17 & V & \nodata\\
2MASS J12115619-7108143 & M3 & $105.28_{-0.75}^{+0.76}$ & 14.530 & 13.021 & 11.828 & 10.09 & 9.51 & 9.24 & \nodata & \nodata\\
2MASS J12203396-7135188 & M4 & 92.36$_{-0.48}^{+0.49}$ & 14.880 & 13.219 & 11.971 & 10.14 & 9.57 & 9.27 & \nodata & \nodata\\
2MASS J12203619-7353027 & M2.5 & 99.40$_{-0.34}^{+0.35}$ & 14.485 & 13.141 & 11.998 & 10.41 & 9.71 & 9.48 & \nodata & \nodata\\
2MASS J12220068-7001041 & M4 & 99.39$_{-0.47}^{+0.47}$ & 15.402 & 14.024 & 12.787 & $\ge$10.40 & $\ge$10.00 & $\ge$9.90 & \nodata & \nodata\\
2MASS J12222238-7137040 & M2 & 94.61$_{-0.24}^{+0.25}$ & 13.769 & 12.530 & 11.425 & 9.90 & 9.27 & 8.99 & \nodata &\nodata\\
2MASS J12224862-7410203 & M5.5 & 93.82$_{-0.78}^{+0.79}$ & 18.688 & 16.175 & 14.698 & 12.26 & 11.69 & 11.29 & P & \nodata\\
2MASS J12255824-7551116 & M4.5 & 98.98$_{-0.63}^{+0.64}$ & 17.324 & 15.366 & 14.034 & 12.02 & 11.49 & 11.19 & \nodata & \nodata\\
2MASS J12324805-7654237 & M3 & 85.17$_{-6.11}^{+7.14}$ & 14.310 & 12.835 & 11.639 & 9.88 & 9.26 & 8.96 & \nodata & \nodata \\
2MASS J12332483-6848553 & M5 & 98.86$_{-0.83}^{+0.85}$ & 17.829 & 15.712 & 14.344 & 12.28 & 11.68 & 11.37 & \nodata & \nodata\\
2MASS J12351540-7043079 & M5 & 92.94$_{-0.56}^{+0.57}$ & 17.853 & 15.701 & 14.317 & 12.18 & 11.66 & 11.33 & \nodata & \nodata\\
2MASS J12421315-6943484 & M5 & 98.85$_{-0.80}^{+0.82}$ & 17.167 & 15.144 & 13.789 & 11.76 & 11.15 & 10.87 & \nodata & \nodata\\
2MASS J12425584-7034207 & M5 & 98.86$_{-0.75}^{+0.77}$ & 18.261 & 15.975 & 14.570 & 12.32 & 11.79 & 11.46 & \nodata & \nodata\\
2MASS J12473611-7031135 & M3 & 99.35$_{-0.30}^{+0.31}$ & 14.628 & 13.269 & 12.119 & 10.50 & 9.86 & 9.71 & \nodata & \nodata\\
\enddata
\tablecomments{a) Spectral types determined  \\ b) Distances have been calculated using the inverse parallax method with zero-point corrections \citep{Lindegren2018}. \\ c) S = suspected spectroscopic binary (M+13); V = resolved binary \citep{Briceno2017}; P = possible photometric binary (this work); C = potential wide separation companion (this work). \\ d) IR excess and/or other evidence indicative of presence of circumstellar disk (M+13 and references therein).}
\end{deluxetable*}

\begin{deluxetable*}{lrrrrrrrr}
\tablecaption{New Wide Separation Companions to $\epsilon$CA Members$^a$}
\tabletypesize{\scriptsize}
\label{table:epschacands}
\tablehead{\colhead{Name} & \colhead{SpT} & \colhead{RA} & \colhead{Dec} & \colhead{$\pi$} & \colhead{G} &  \colhead{G$-$G$_{RP}$} & \colhead{PMRA} & \colhead{PMDec}  \\ \colhead{} & \colhead{} &\colhead{(deg)} & \colhead{(deg)} & \colhead{(mas)} & \colhead{(mag)} & \colhead{(mag)} & \colhead{(mas/yr)} & \colhead{(mas/yr)}}
\startdata
2MASS J11550485-7919108 & M3 & 178.7692 & -79.3198 & 9.886$\pm$0.058 & 14.818 & 1.35 & -41.18$\pm$0.13 & -4.34$\pm$0.09\\
2MASS J11550336-7919147 & M6 & 178.7628 & -79.3208 & 9.82$\pm$0.53 & 19.92 & 1.74 & -39.74$\pm$1.22 & -4.66$\pm$0.68 \\
\hline
HD 104467 & G3 & 180.4120 & -78.9881 & 10.14$\pm$0.17 & 8.428 & 0.515 & -41.11$\pm$0.26 & -5.41$\pm$0.26\\
2MASS J12011981-7859057 & M5 & 180.3316 & -78.9849 & 9.794$\pm$0.075 & 15.179 & 1.3562 & -41.99$\pm$0.12 & -5.35$\pm$0.10\\
\hline
HD 105923 & G8 & 182.9084 & -71.1767 & 9.400$\pm$0.036 & 8.837 & 0.537 & -38.720$\pm$0.060 & -7.42$\pm$0.05\\
2MASS J12115619-7108143 & M3 & 182.9836 & -71.1374 & 9.44$\pm$0.68 & 13.021 & 1.1928 & -38.64$\pm$0.11 & -8.11$\pm$0.09 \\
\hline
\enddata
\tablecomments{a) Candidate wide-separation comoving systems are listed as pairs, with the previously identified $\epsilon$CA member listed first and its candidate wide-separation companion listed second.}
\end{deluxetable*}

\begin{deluxetable*}{lrrrrrrrrrrrrr}
\tablecaption{Bona Fide $\epsilon$CA Members: Heliocentric Positions and Velocities}
\tabletypesize{\scriptsize}
\label{table:uvw}
\tablehead{\colhead{Name} & \colhead{X} & \colhead{$\sigma$} & \colhead{Y} & \colhead{$\sigma$} & \colhead{Z} & \colhead{$\sigma$} & \colhead{U} & \colhead{$\sigma$} & \colhead{V} & \colhead{$\sigma$} & \colhead{W} & \colhead{$\sigma$}\\ \colhead{} & \colhead{} & \colhead{}  & \colhead{} & \colhead{[pc]} &  \colhead{}  & \colhead{} & \colhead{} & \colhead{} & \colhead{} & \colhead{[km s$^{-1}$]} & \colhead{} & \colhead{}  & \colhead{} }
\startdata
CP-68 1388 & 44.89 & 0.15 & -105.20 & 0.35 & -17.209 & 0.057 & -10.86 & 0.39 & -20.69 & 0.91 & -8.72 & 0.15\\ 
2MASS J11183572-7935548A & 43.049 & 0.605 & -78.8 & 1.1 & -28.32 & 0.40 & -7.80 & 0.77 & -23.6 & 1.3 & -10.37 & 0.50 \\ 
2MASS J11183572-7935548B & 42.5 & 1.6 & -77.9 & 2.9 & -28.0 & 1.0 & \nodata & \nodata & \nodata & \nodata & \nodata & \nodata\\ 
RX J1147.7-7842 & 47.86 & 0.19 & -83.82 & 0.33 & -28.09 & 0.11 & -9.54 & 0.44 & -20.99 & 0.75 & -11.23 & 0.25\\ 
RX J1149.8-7850 & 47.89 & 0.12 & -83.46 & 0.21 & -28.230 & 0.071 & -10.87 & 0.62 & -18.8 & 1.1 & -10.39 & 0.37\\ 
RX 1150.9-7411 & 44.08 & 0.75 & -80.3 & 1.4 & -19.16 & 0.33 & -7.88 & 0.64 & -19.3 & 1.0 & -10.57 & 0.29\\ 
2MASS J11550485-7919108 & 48.51 & 0.29 & -83.23 & 0.49 & -28.98 & 0.17 & \nodata & \nodata & \nodata & \nodata & \nodata & \nodata\\ 
T Cha & 52.712 & 0.474 & -90.06 & 0.81 & -31.43 & 0.28 & -11.98 & 0.65 & -19.4 & 1.1 & -13.03 & 0.38\\ 
RX J1158.5-7754B & 49.75 & 0.18 & -85.89 & 0.31 & -27.220 & 0.098 & -10.45 & 0.97 & -18.5 & 1.7 & -10.00 & 0.53\\ 
RX J1158.5-7754A & 50.50 & 0.19 & -87.18 & 0.32 & -27.62 & 0.10 & -10.06 & 0.88 & -18.3 & 1.5 & -13.71 & 0.48\\ 
HD 104036 & 50.25 & 0.20 & -86.80 & 0.35 & -27.34 & 0.11 & -11.49 & 0.25 & -18.26 & 0.41 & -11.07 & 0.14 \\ 
CXOU  J115908.2-781232 & 51.08 & 0.33 & -87.83 & 0.57 & -28.40 & 0.19 & -9.45 & 0.16 & -20.27 & 0.18 & -10.398 & 0.076\\ 
RX J1159.7-7601 & 47.87 & 0.11 & -83.77 & 0.19 & -23.116 & 0.053 & -10.4 & 1.8 & -18.7 & 3.1 & -9.65 & 0.86\\ 
HD 104237A & 52.25 & 0.33 & -89.68 & 0.56 & -28.96 & 0.18 & -10.85 & 0.23 & -19.15 & 0.33 & -10.60 & 0.12\\ 
HD 104237D & 48.79 & 0.30 & -83.73 & 0.52 & -27.04 & 0.17 & \nodata & \nodata & \nodata & \nodata & \nodata & \nodata\\ 
HD 104237E & 49.23 & 0.17 & -84.49 & 0.29 & -27.290 & 0.093 & \nodata & \nodata & \nodata & \nodata & \nodata & \nodata\\ 
2MASS J12005517-7820296 & 49.72 & 0.42 & -85.09 & 0.71 & -27.76 & 0.23 & -11.92 & 0.65 & -16.9 & 1.1 & -9.01 & 0.36\\ 
HD 104467 & 47.74 & 0.77 & -81.1 & 1.3 & -27.63 & 0.45 & -10.39 & 0.55 & -18.32 & 0.81 & -9.65 & 0.30\\ 
2MASS J12014343-7835472 & 50.75 & 0.56 & -86.52 & 0.95 & -28.71 & 0.32 & -8.02 & 0.36 & -24.72 & 0.51 & -12.37 & 0.19\\ 
USNO-B 120144.7-781926 & 49.26 & 0.32 & -84.16 & 0.54 & -27.42 & 0.18 & -10.02 & 0.55 & -20.34 & 0.92 & -10.61 & 0.30\\ 
CXOU J120152.8-781840 & 49.69 & 0.34 & -84.88 & 0.59 & -27.62 & 0.19 & -9.02 & 0.55 & -21.44 & 0.92 & -11.36 & 0.30\\ 
RX J1202.1-7853 & 48.37 & 0.22 & -82.22 & 0.38 & -27.80 & 0.13 & -10.18 & 0.13 & -23.05 & 0.17 & -10.661 & 0.067\\ 
RX  J1202.8-7718 & 50.37 & 0.18 & -86.55 & 0.31 & -26.265 & 0.094 & -9.85 & 0.30 & -19.89 & 0.50 & -10.11 & 0.16\\ 
RX J1204.6-7731 & 48.92 & 0.17 & -83.61 & 0.28 & -25.761 & 0.087 & -11.98 & 0.98 & -16.67 & 1.67 & -9.11 & 0.52\\ 
RX J1207.7-7953 & 48.77 & 0.15 & -81.38 & 0.25 & -29.325 & 0.089 & -9.88 & 0.35 & -20.29 & 0.58 & -10.92 & 0.21\\ 
HD 105923 & 51.83 & 0.20 & -90.86 & 0.35 & -15.736 & 0.060 & -9.64 & 0.52 & -20.67 & 0.91 & -8.76 & 0.16\\ 
RX J1216.8-7753 & 50.43 & 0.20 & -83.74 & 0.33 & -26.46 & 0.10 & -9.7 & 1.0 & -19.5 & 1.6 & -10.39 & 0.52\\ 
RX J1219.7-7403 & 50.50 & 0.14 & -84.98 & 0.23 & -19.808 & 0.053 & -9.71 & 0.95 & -20.02 & 1.59 & -9.40 & 0.37\\ 
2MASS J12210499-7116493 & 49.61 & 0.12 & -84.31 & 0.20 & -14.713 & 0.035 & -10.5 & 1.3 & -18.16 & 2.16 & -8.38 & 0.38\\ 
RX J1239.4-7502 & 53.60 & 0.15 & -85.33 & 0.24 & -21.772 & 0.060 & -9.3 & 1.5 & -19.8 & 2.3 & -9.66 & 0.59\\ 
CD-69 1055 & 51.63 & 0.14 & -77.94 & 0.22 & -12.501 & 0.034 & -9.8 & 1.0 & -18.7 & 1.6 & -8.28 & 0.25\\ 
MP Mus & 56.85 & 0.17 & -79.36 & 0.24 & -11.857 & 0.036 & -9.13 & 0.13 & -19.64 & 0.17 & -8.566 & 0.043\\
\enddata
\end{deluxetable*}

\begin{deluxetable*}{lrrrr}
\tablecaption{Mean and Median Heliocentric Positions and Velocities of $\epsilon$CA Members}
\label{table:epschameans}
\tablehead{\colhead{} & \colhead{Mean$^a$} & \colhead{Median} & \colhead{$\sigma^b$} }
\startdata
Distance [pc] & 100.99 & 100.81 & 4.62\\
\hline
X [pc] & 49.660 & 49.650 & 2.854 \\
Y [pc] & -84.328 & -84.232 & 5.185 \\
Z [pc] & -18.905 & -27.378 & 5.143 \\
\hline
U [km s$^{-1}$] & -9.847 & -9.954 & 1.089 \\
V [km s$^{-1}$] & -20.667 & -19.557 & 1.788 \\
W [km s$^{-1}$] & -9.682 & -10.382 & 1.305 \\
\enddata
\tablecomments{a) Weighted means and medians as calculated from values listed in Table~\ref{table:uvw}. \\ b) Standard deviation of the corresponding weighted mean. }
\end{deluxetable*}

\end{document}